\documentclass[namedreferences,hyperref,optionalrh,solaromanenum]{spr-sola}

\usepackage{graphicx}                    
\usepackage{color}                    
\usepackage{amsmath}
\usepackage{amssymb}
\usepackage{array}
\usepackage{multirow}
\usepackage{color}
\definecolor{goldenyellow}{rgb}{1.0, 0.87, 0.0}
\definecolor{orange}{rgb}{1.0, 0.5, 0.0}
\usepackage{lmodern}

\chardef\us=`\_

\begin{document}

\begin{frontmatter}

\title{An analysis of the solar differential rotation in solar cycle No. 19 (1954-1964) determined using Kanzelh{\"o}he sunspot group positions}

%
\address[id=aff1]{University of Rijeka, Faculty of Physics, Radmile Matej\v{c}i\'{c} 2, 51000 Rijeka, Croatia}
\address[id=aff2]{Zagreb Astronomical Observatory, Opati\v{c}ka 22, 10000 Zagreb, Croatia}
\address[id=aff3]{Kanzelh{\"o}he Observatory for Solar and Environmental Research, University of Graz, Kanzelh{\"o}he 19, 9521 Treffen am Ossiacher See, Austria}
\address[id=aff4]{Hvar Observatory, Faculty of Geodesy, University of Zagreb, Ka\v{c}i\'{c}eva 26, 10000 Zagreb, Croatia}
\address[id=aff5]{Institute of Physics, University of Graz, Universit{\"a}tsplatz 5, 8010 Graz, Austria }

\author[addressref={aff1},corref,email={ipoljancic@phy.uniri.hr}]{\inits{I.}\fnm{Ivana}~\snm{Poljan\v{c}i\'{c} Beljan}\orcid{0000-0002-4887-815X}}
\author[addressref={aff1}, email={}]{\inits{L.}\fnm{Luka}~\snm{\v{S}ibenik}\orcid{}}
\author[addressref={aff1}, email={}]{\inits{T.}\fnm{Tomislav}~\snm{Jurki\'{c}}\orcid{0000-0002-4993-2939}}
\author[addressref={aff1}, email={}]{\inits{K.}\fnm{Klaudija}~\snm{Lon\v{c}ari\'{c}}\orcid{0000-0003-4794-3682}}
\author[addressref={aff1}, email={}]{\inits{R.}\fnm{Rajka}~\snm{Jurdana-\v{S}epi\'{c}}\orcid{0000-0001-8361-4494}}
\author[addressref={aff2}, email={}]{\inits{D.}\fnm{Damir}~\snm{Hr\v{z}ina}\orcid{}}
\author[addressref={aff3}, email={}]{\inits{W.}\fnm{Werner}~\snm{P{\"o}tzi}\orcid{0000-0003-4811-0543}}
\author[addressref={aff4}, email={}]
{\inits{R.}\fnm{Roman}~\snm{Braj\v{s}a}\orcid{0009-0008-3048-8138}}
\author[addressref={aff3,aff5}, email={}]{\inits{A. M.}\fnm{Astrid M.}~\snm{Veronig}\orcid{0000-0003-2073-002X}}
\author[addressref={aff5}, email={}]{\inits{A.}\fnm{Arnold}~\snm{Hanslmeier}\orcid{0000-0002-7282-5007}}

%
\runningauthor{I. Poljan\v{c}i\'{c} Beljan et al.}
\runningtitle{Photospheric differential rotation analysis for solar cycle No. 19}

\begin{abstract}
We study solar differential rotation for solar cycle No. 19 (1954–1964) by tracing sunspot groups on the sunspot drawings of Kanzelh{\"o}he Observatory for Solar and Environmental Research (KSO). Our aim is to extend previous differential rotation (DR) analysis from the KSO data (1964-2016) to the years prior to 1964 to create a catalog of sunspot group positions and photospheric DR parameters from KSO sunspot drawings and white light images. Synodic angular rotation velocities were first determined using the daily shift (DS) and robust linear least-squares fit (rLSQ) methods, then converted to sidereal velocities, and subsequently used to derive solar DR parameters. We compare the DR parameters obtained from different sources and analyse the north–south asymmetry of rotation for solar cycle No. 19. It has been shown that our results for the equatorial rotation velocity (parameter $A$) and the gradient of DR (parameter $B$) coincide with earlier results from the KSO data (performed with a different method), as well as with results from the Kodaikanal Solar Observatory (KoSO) and the Yunnan Observatories (YNAO). In contrast, the values of parameter $A$ from three different earlier studies based on the Greenwich Photoheliographic Results (GPR) exhibit statistically significant differences when compared to the values of parameter $A$ derived from KSO, KoSO and YNAO. These findings suggest that the GPR data have the largest inconsistency compared to the other three data sources, highlighting the need for further analysis to identify the causes of these discrepancies. The analysis of the north-south asymmetry in the solar rotation profile using two different methods shows that the DR parameters of the hemispheres coincide, indicating a rotational symmetry around the equator. This is consistent with previous results from KSO and YNAO data. However, all sources indicate slightly higher equatorial rotation velocities in the southern hemisphere.
\end{abstract}

%

\end{frontmatter}


\section{Introduction}\label{S-Introduction} 
Historical sunspot observations are invaluable to solar physicists because recovering these records from historical libraries and archives is key to understanding the evolution of the Sun over the centuries
\citep{vaquero2009}. \citet{arlt2020} discussed the value of historical sunspot drawings derived from naked eye and telescopic observations. They pointed out that the positional information as well as the morphological information accessible from the drawings are essential for performing improved estimation of sunspot number, average group locations, tilt angles of the groups, polarity separation of groups, sunspot growth and decay rates, solar rotation, north–south asymmetry of rotation and activity, and the relationship between solar rotation and activity. Experimental results on the long-term variability of these quantities over centuries serve as important information for understanding the solar dynamo \citep[e.g.][]{charbonneau2010}.
There have been many observers from telescopic era of sunspot observations which enabled positional data of sunspots and sunspot groups (we highlight some): Harriot \citep{vokhmyanin2020}, Galilei \citep{vokhmyanin_galileo2018}, Gassendi \citep{vokhmyanin_gassendi2018}, Scheiner \citep{arlt2016}, Marcgraf \citep{vaquero2011}, Hevelius \citep{carrasco2019}, observations at Paris Observatory \citep{vaquero2015}, Kirch \citep{neuhauser2018}, Becker \citep{neuhauser2015}, Wargentin \citep{arlt2018}, Staudacher \citep{arlt_staudacher2009}, Horrebow \citep{karoff2019}, Hamilton \citep{arlt_hamilton2009}, Schwabe \citep{arlt2013}, Spörer \citep{diercke2015}, and the famous Greenwich Photoheliographic Results (GPR, 1874-1976) of Royal Greenwich Observatory, continued by Solar Optical Observing Network/United States Air Force/National Oceanic and Atmospheric Administration data set (SOON/USAF/NOAA\footnote{\url{https://solarscience.msfc.nasa.gov/greenwch.shtml}}, 1976-2016), as shown in Figures 26 and 27 of \citet{arlt2020}. 
In addition to the SOON/USAF/NOAA data, the Debrecen Photoheliographic Data (DPD\footnote{\url{http://fenyi.solarobs.epss.hun-ren.hu/DPD/}}, 1976-2018) sunspot catalog also represents a continuation of the GPR \citep{baranyi2016, gyori2017}. Unfortunately, the SOON/USAF/NOAA and DPD catalogs no longer provide additional data as of 2018. Previous analyses \citep{poljancic2017, poljancic2022} showed that the Kanzelh{\"o}he Observatory for Solar and Environmental Research (KSO) dataset is in good agreement with DPD and GPR, making it suitable for investigating long-term variabilities in the solar rotation profile. KSO is therefore a good candidate to fill the gaps left by the discontinuation of DPD and SOON/USAF/NOAA.

The database of sunspot group positions, photospheric differential rotation (DR) parameters, their changes over time, north–south rotational asymmetry, and the dependence of DR on solar activity have already been determined from the KSO sunspot drawings and white light images for the timespan 1964–2016 \citep{poljancic2017, poljancic2022}. We aim to extend this research to a longer period by expanding the database of sunspot group positions from KSO to include the years before 1964 and then the years after 2016. Since the first year in which sunspot drawings appear at KSO is 1944, the 53 years processed so far (1964–2016) will be extended by another 30 years available. The completion of the user-friendly catalog of KSO sunspot group positions, rotation velocities and differential rotation parameters is therefore very important for the further long-term analysis of photospheric differential rotation, considering the discontinuation of DPD and SOON/USAF/NOAA.

KSO is the only solar observatory in Austria for solar and environmental research and is part of the Institute of Physics at the University of Graz. It provides daily multispectral synoptic observations of the Sun, and the complete archive of sunspot drawings since 1944 has been digitized\footnote{\url{http://cesar.kso.ac.at/synoptic/draw_years.php}}. More details about the KSO sunspot drawings can be found in \citet{otruba2006} and \citet{potzi2016, potzi2021}.
This paper presents the results of solar differential rotation for solar cycle No. 19 (1954–1964) derived from tracing sunspot groups on KSO sunspot drawings. In addition to determining the positions of the sunspot groups and the photospheric solar differential rotation, we compare the differential rotation parameters obtained from different sources for solar cycle No. 19. We also analyse the north-south asymmetry in the solar rotation profile during this cycle. 

Solar cycle No. 19 is particularly interesting because its peak coincides with the modern Gleissberg maximum. The extension of the KSO dataset allows us to validate the results of \citet{poljancic2022}, which indicate a phase shift between the activity reversal (modern Gleissberg maximum) and the reversal of the equatorial rotation velocity (beginning of the acceleration in the early 1990s). In addition, solar cycle No. 19 exhibits an extreme north-south asymmetry of activity, with the northern hemisphere showing an excess of activity for 69 months (50\% of the cycle) between 1954 and 1964, while the southern hemisphere only had an excess of activity for 11 months \citep{temmer2006, veronig2021}.


\section{Methods}\label{S-Methods} 

KSO sunspot drawings are recorded using a refractor telescope ($d / f = 110 / 1650$ mm) with a projection system that magnifies the solar disk image to a diameter of 25 cm. These drawings are uploaded online on a daily basis, depending on favorable weather, which typically permits around 300 observation days per year. A comprehensive overview of the KSO’s solar monitoring program, data products, and sunspot drawings is provided by \citet{otruba2006} and \citet{potzi2016, potzi2021}. For the analysis, we employed an interactive (int) procedure for the determination of tracer’s positions in full disk solar images using the Sungrabber software package \citep{hrzina2007}. Sunspot group centers were estimated visually, with emphasis placed on prominent umbra and penumbra regions. Thus, a visual area-weighted method was applied, in which assessor was responsible for consistently evaluating the position based on three guiding steps: (1) the position must lie within the group, (2) the position tends toward the region containing more and larger spots, (3) in the chronologically next image, the observer estimates the position based on the previous two criteria while also taking into account the previously estimated positions (this condition ensures the same group center position when there are no significant morphological changes, and provides an accurate reference in cases of major morphological changes).

\begin{figure}
   \centering
   \resizebox{12cm}{!}{\includegraphics[width=1.2\textwidth,clip=]{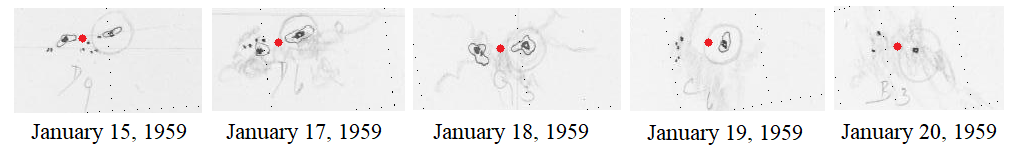}}\     \caption{Example of the determination of the center of sunspot group 19134. The red dot indicates the determined position of the sunspot group's center.}
     \label{Fig1}
\end{figure}

Fig.~\ref{Fig1} shows several examples of center position determinations for sunspot group 19134 during consecutive days, January 15–20, 1959. When the area of the leading spot significantly exceeds that of the following pore(s), the area-weighted center shifts toward the leading part — see Fig.~\ref{Fig1}, January 19, 1959. In cases where the pores continue to decay and become even smaller in the following days, the area-weighted center shifts further toward the leading spot, almost overlapping with it. Naturally, if we have a bipolar symmetric sunspot group, as in Fig.~\ref{Fig1} for January 18, 1959, the position is located in the middle. 

Clearly, the visual area-weighted procedure may be subjective, assessor may occasionally over- or underestimate the influence of a dominant spot or different assessors may interpret the relative sizes of the individual spots within the group differently, leading to variations in center position estimates. For this reason, before beginning the position determinations for our first paper \citep{poljancic2017}, we conducted a test: two independent assessors determined the positions of 10 sunspot groups (both H and J and complex groups). The differences between their measurements were shown to be approximately 0.1° per day \citep{poljancic2010}, which is smaller than the discrepancies between different observatories \citep{poljancic2011}.

Furthermore, the consistency in applying the same method as in our previously published results using the processed portion of KSO data so far (1964–2016) \citep{poljancic2017,ruzdjak2018,poljancic2022} justifies the continued use of the visual method. This consistency is also important for the long-term goal of creating a comprehensive KSO catalog (1944 – today) of sunspot group positions. Therefore, our conclusion is that, although the use of the visual area-weighted procedure may occasionally lead to over- or underestimating the influence of a dominant spot, as long as a consistent and agreed-upon procedure is applied throughout, the results should remain reliable and trustworthy.

\begin{figure}
   \centering
   \resizebox{6cm}{!}{\includegraphics[width=0.7\textwidth,clip=]{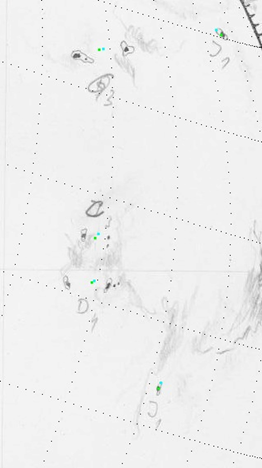}}\     \caption{Part of the KSO sunspot drawing from November 26, 1958, with marked KSO (green) and GPR (blue) positions of the sunspot groups, that are measured only 8 minutes apart.}
     \label{Fig2}
\end{figure}

The Sungrabber software enables users to load the heliographic positions of the sunspot groups from GPR or DPD together with the sunspot group numbers from Greenwich or NOAA/USAF. Therefore, using the “Compare” option in Sungrabber, sunspot groups in this study were identified by the corresponding Greenwich sunspot group numbers, along with Greenwich positions for reference. Additionally, in KSO sunspot drawings, when two nearby sunspot groups appear, their separation is almost always clearly indicated by line, with assigned Z{\"u}rich classification numbers labeled directly on the images. In cases where uncertainty still remained, we consulted the GPR Sunspot Catalogue\footnote{\url{http://fenyi.solarobs.epss.hun-ren.hu/GPR/1948/index.html}} \citep{baranyi2016}, and compared their full-disk images with the ones from KSO.

The GPR positions do not always align precisely with KSO ones, as the observations were not taken at the exact same time of day, resulting in slight positional shifts. However, on days when the observation times were nearly identical, the determined positions often match closely (Fig~\ref{Fig2}).

The current version of the Sungrabber software\footnote{\url{https://zvjezdarnica.hr/sungrabber/sungrabb.html}} has been significantly improved compared to the former version from 2007, especially in the part of the determination of the Solar limb and the daily motion that is semi–automated. Loading and handling solar drawing files is also simplified, significantly accelerating processing time.

\begin{figure}
   \centering
   \resizebox{10cm}{!}{\includegraphics[bb = -85 15 605 355]{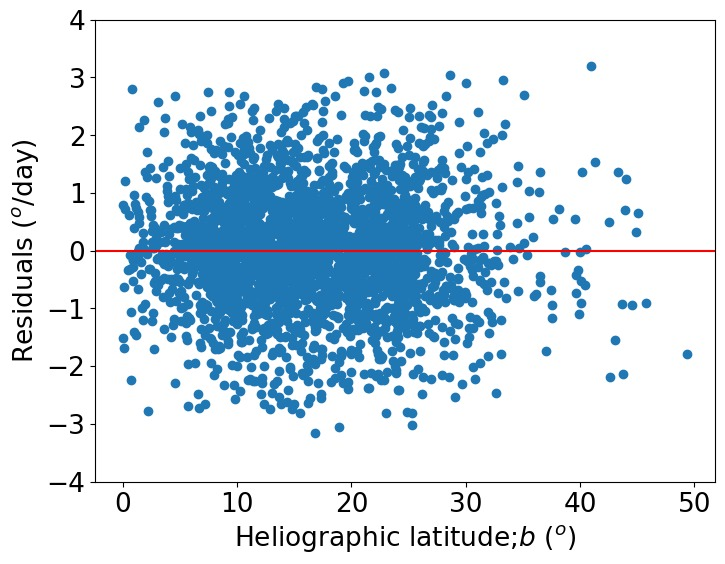}}\     \caption{Residuals of the DS sidereal rotation velocities against the explanatory variable - heliographic latitude $b$.}
     \label{Fig3}
\end{figure}

\begin{figure}
   \centering
   \resizebox{10cm}{!}{\includegraphics[bb = -128 15 555 265]{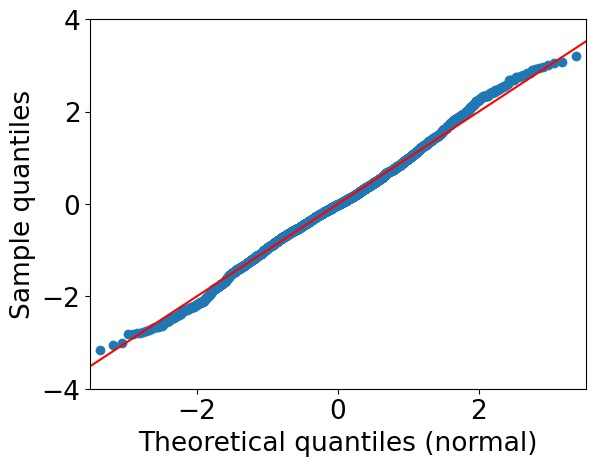}}\     \caption{A Q–Q plot of a sample of data (vertical axis) versus a Normal distribution (horizontal axis).}
     \label{Fig4}
\end{figure}

Rotation velocities were calculated using two methods. The first method is called the daily-shift method (DS), deriving the synodic rotation velocity $\omega_{syn}=\Delta CMD/\Delta t$ from daily changes in the Central Meridian Distance ($CMD$) and elapsed time $\Delta t$. The second method makes use of a robust linear least-squares fit (rLSQ) applied to the $CMD(t)$ function for each individual tracer (sunspot group). In the second method, the synodic rotation velocity is represented by the slope of the fit $CMD(t)$ versus $t$. Fitting and determining the synodic velocity has been performed for each sunspot group with at least three measured positions. To account for occasional outliers in the data, caused by misidentifications or other factors, a robust fitting approach was applied. This method employs an iteratively reweighted least-squares fit using Huber’s $t$ weighting function \citep{huber1981}. 

For a given individual sunspot group, the DS method yields multiple rotation velocities, while the rLSQ method produces a single velocity. As a result, the number of calculated velocities is several times lower for the rLSQ method compared to the DS method. To harmonize the number of derived velocities, we calculated mean DS velocities, assigning a single rotation velocity value to each group. Consequently, the number of data points $(N_\mathrm{vl})$ listed for the DS method and rLSQ method in Table 1 is almost the same, DS method slightly exceeding the latter. This discrepancy arises primarily from the cases where only two positions were measured for a group—sufficient to calculate a DS velocity but insufficient for rLSQ, which requires at least three data points to perform a linear fit and determine a rotation velocity. 

After calculation, velocities were converted from synodic to sidereal values \citep{rosa1995, brajsa2002solphys, skokic2014}. To reduce errors, the data were limited to $CMD$ values within $\pm 58^{\circ} $ to avoid solar limb effects and to a sidereal rotation velocity range of $11$–$17^{\circ}$ per day, which eliminates significant errors such as those arising from sunspot group misidentification. The former filter is based on typical values used in various analyses of solar rotation, while the latter range is specifically set to remove irregularities in the distribution of the residuals of the sidereal rotation velocity in our data, namely heavy tails.

In the daily-shift method, each calculated velocity was assigned to the latitude and time of the first recorded position measurement \citep{olemskoy2005}. As the applied interactive procedure is the same as the one used in \citet{poljancic2017}, additional information on the instrumentation and measurement processes along with more in-depth descriptions of the methods for determining heliographic positions and rotation velocities is available in Sects. 2 – 4 of the same paper. Finally, the DR law 
   \begin{equation}
    \label{eq1}
      \omega(b) = A+B \sin^2b
   \end{equation}
was used to obtain the DR parameters $A$ and $B$, where $\omega$ is the sidereal velocity and $b$ is the heliographic latitude. The Python function numpy.polyfit() was used for the calculations. Fitting was performed for both methods (DS and rLSQ), the whole cycle and separately for the northern and southern hemispheres of the Sun. 

\begin{table*}[ht]
\caption{Results of fittings of the KSO data and data used by other authors (solar cycle No. 19) to rotational velocity profile according to Eq. (\ref{eq1}). The number of calculated rotation velocities is denoted as $N_{\mathrm{vl}}$, with applied $\pm58^{\circ}$ $CMD$ filter and 11–17$^{\circ}$day$^{-1}$ velocity filter. N+S represents both hemispheres combined, while N and S refer to the northern and southern hemispheres, respectively. References: $^a$present paper; $^b$\citet{lustig1983}; $^c$\citet{balthasarvazquez1986}; $^d$\citet{pulkkinen1998}; 
$^e$\citet{javaraiah2005}; $^f$\citet{jha2021}; $^g$\citet{luo2021}}
\label{table1}
\centering
\begin{tabular}{llllll}
\hline
Row & Data set / Method / Hemisphere & $A$ (deg/day) & $B$ (deg/day) & $N_\mathrm{vl}$ \\
\hline
1   & KSO$^a$ / int, DS 	   /   N+S   & $14.350\pm0.021$  & $-2.138\pm0.159$  & 2740 \\
2   & KSO$^a$ / int, DS   /  N     & $14.344\pm0.027$  & $-2.052\pm0.196$	& 1594 \\
3   & KSO$^a$ / int, DS	    /   S     & $14.362\pm0.033$ & $-2.296\pm0.269$	& 1146 \\
4   & KSO$^a$ / int, rLSQ   /   N+S   & $14.392\pm0.019$  & $-2.696\pm0.146$	& 2266 \\
5   & KSO$^a$ / int, rLSQ   /   N     & $14.359\pm0.024$  & $-2.551\pm0.182$	& 1326 \\
6   & KSO$^a$ / int, rLSQ   /   S     & $14.441\pm0.029$  & $-2.922\pm0.242$	& 940 \\
7   & KSO$^b$  /Stonyhurst, LSQ    /  N & $14.36\pm0.02$    & $-2.76\pm0.12$  &  -\\
8   & KSO$^b$  / Stonyhurst, LSQ   /  S	& $14.38\pm0.02$    & $-2.52\pm0.17$  & - \\
9   & GPR$^c$  / DS  /  N+S	& $14.51\pm0.02$  & $-2.84\pm0.11$  & 14733 \\
10  & GPR$^d$  / DS  /  N+S	& $14.512\pm0.013$  & $-2.798\pm0.105$  & 17803 \\
11  & GPR$^e$  / DS   /  N+S	& $14.486\pm0.009$   & $-2.678\pm0.077$  & 18929 \\
12  & KoSO$^f$ / DS/ N+S	& $14.370\pm0.014$   & $-2.780\pm0.115$    & - \\
13  & YNAO$^{g}$ / Method\tabnote{Explanation of the method: $\omega(b)=14.18^{\circ}$day$^{-1} + m^{\circ}/n$, where $n$ is interval of time - number of mean solar days, $m^{\circ}$ the change in heliographic longitude, $m^{\circ}/n$ diurnal angle motion relative to central meridian and $14.18^{\circ}$day$^{-1}$ the sidereal rotation velocity of the central meridian.}  ~/  N+S	& $14.335\pm0.021$   & $-2.394\pm0.204$  & - \\
14  & YNAO$^{g}$ / Method$^1$ /  N	& $14.328\pm0.026$ & $-2.257\pm0.226$  & - \\
15  & YNAO$^{g}$ / Method$^1$ /  S	& $14.371\pm0.037$ & $-3.141\pm0.504$  & - \\
\hline
\end{tabular}
\end{table*}

To compare different rotation profiles, we applied a statistical criterion like the one in \citet{brajsa1995}. The difference between two solar rotation rates was considered statistically significant if it exceeded three times the sum of their standard errors. Conversely, the difference between two results (e.g. the parameters $A_1$ and $A_2$) can be considered statistically insignificant if it is smaller than the sum of their individual 1$\sigma$ uncertainties
   \begin{equation}
    \label{eq2}
      \Delta A = A_1 - A_2 < 1\sigma(A_1)+1\sigma(A_2)
   \end{equation}
Here $\sigma$ stands for the standard error. All values that fulfill Eq. (\ref{eq2}) are considered statistically equivalent.

To correctly interpret and evaluate results, it is necessary to check the assumptions of the least-squares method (i.e., homoscedasticity and normality of the residuals of the sidereal rotation velocity)\footnote{\url{https://devopedia.org/linear-regression}}. Homoscedasticity, or the constant variance of the residuals, was verified by plotting the residuals against the heliographic latitude ($b$) on a scatter plot\footnote{\url{https://statsnotebook.io/blog/analysis/linearity_homoscedasticity/}}, see Fig.~\ref{Fig3}.

The residuals were calculated as the difference $y_\mathrm{exp}-y_\mathrm{model}$, where $y_\mathrm{exp}$ represents "measured" sidereal rotation velocities, and $y_\mathrm{model}=A+B\cdot x$ is the sidereal rotation velocity determined by inserting the values of the parameters $A$ and $B$ from Table~\ref{table1}, row 1 for the DS method and row 4 for the rLSQ method. Here, $x$ represents $\sin^2b$, where $b$ is heliographic latitude corresponding to particular $y_\mathrm{exp}$.

Normality of the residuals was examined using a Quantile-Quantile (Q-Q) plot, checking for large deviations from the $45^{\circ}$ reference line, which represents a normal distribution\footnote{\url{https://statisticsbyjim.com/graphs/qq-plot/It}}, see Fig.~\ref{Fig4}. Although the distribution of the residuals is not exactly normal, according to the theorem of asymptotic normality, as the sample size becomes sufficiently large, the distribution of the estimators will approach a normal distribution\footnote{\url{https://www.statisticshowto.com/asymptotic-normality/}}. As a result, a sample size of 25 was set as the minimum for interval construction and the conduction of statistical tests. Images for checking the assumptions of the least-squares method are shown only for the DS method.


\section{Results and discussion}\label{S-Results} 

\subsection{Differential rotation from KSO data for cycle No. 19}\label{S-DR} 

The results of photospheric solar differential rotation during the solar cycle No. 19 (1954–1964), derived by tracing sunspot groups on KSO sunspot drawings, are presented. In Table~\ref{table1} (rows 1-6) we list the results of fittings to Eq. (\ref{eq1}) for the whole solar cycle No. 19 of KSO data (1954 – 1964): for both methods (DS and rLSQ), the analysis was performed for both hemispheres combined (N+S) as well as separately for the northern (N) and southern (S) hemispheres. 

\begin{figure}   
\centerline{\hspace*{0.015\textwidth}
         \includegraphics[width=1.015\textwidth,clip=]{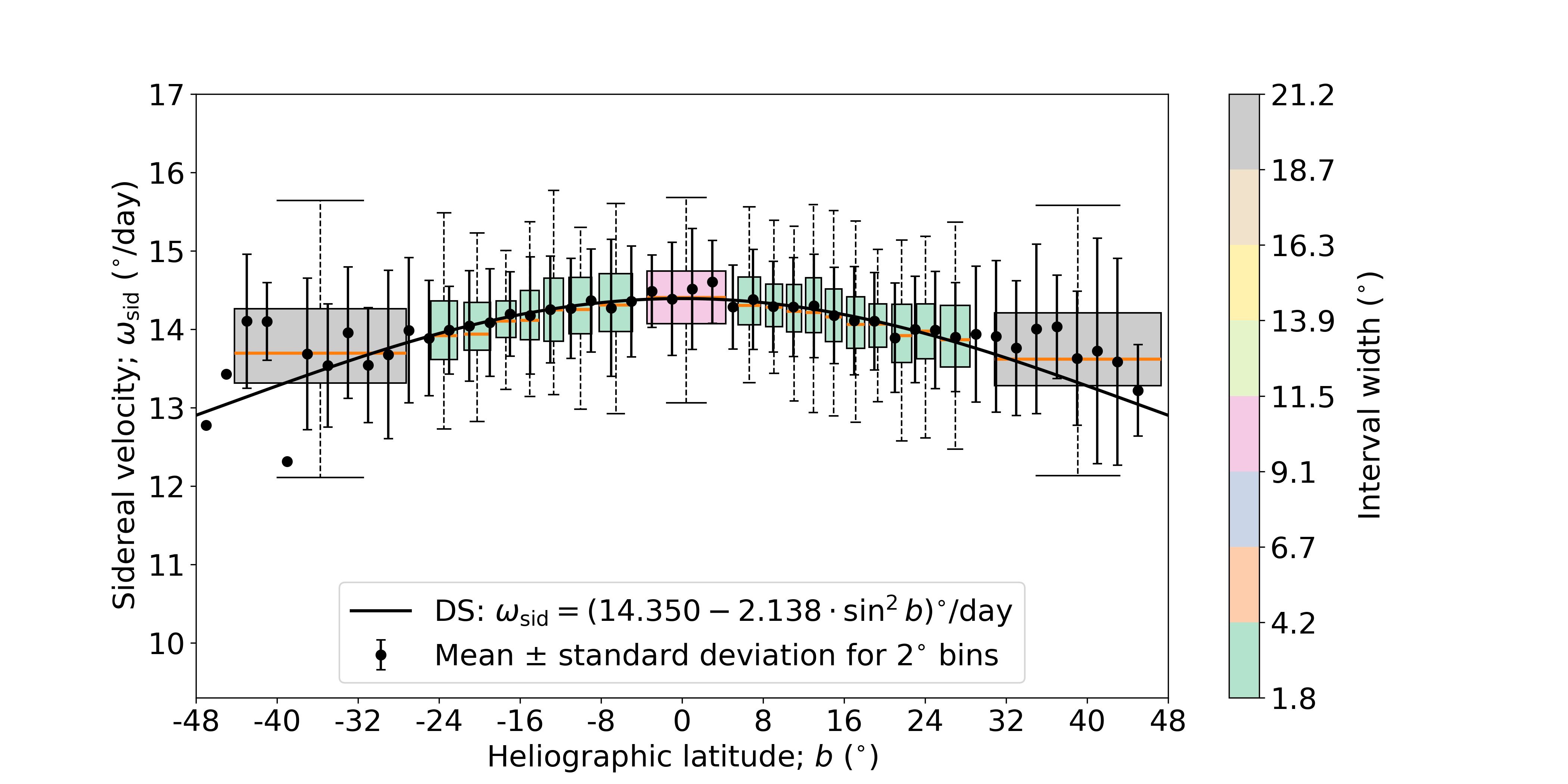}
         \hspace*{-0.03\textwidth}
        }
\vspace{-0.36\textwidth}   
\centerline{\Large \bf     
\hspace{0.0\textwidth} \color{black}{(a)}
   \hfill}
\vspace{0.31\textwidth}    
\centerline{\hspace*{0.025\textwidth}
         \includegraphics[width=1.015\textwidth,clip=]{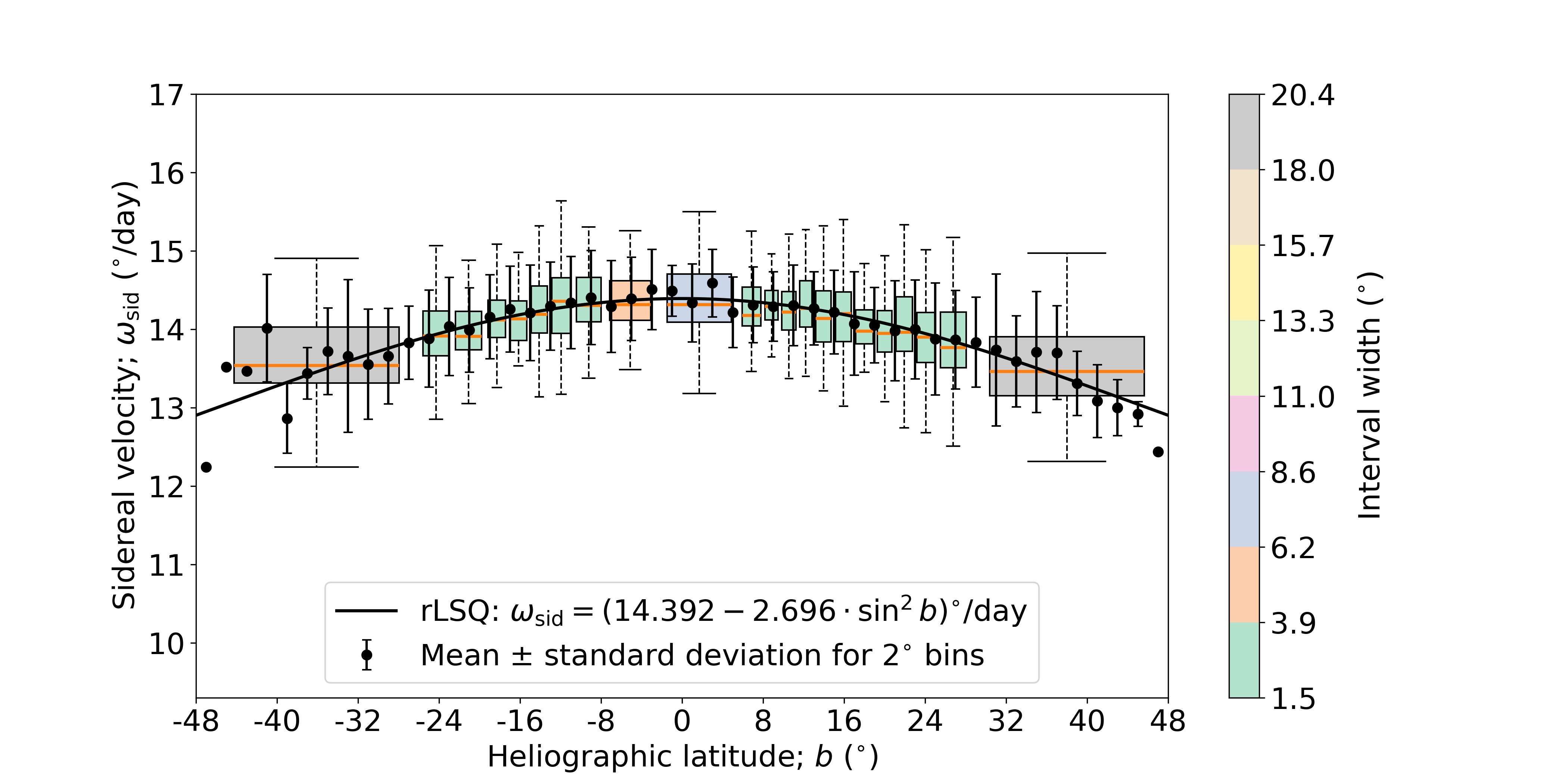}
         \hspace*{-0.03\textwidth}
        }
\vspace{-0.35\textwidth}   
\centerline{\Large \bf     
\hspace{0.0 \textwidth} \color{black}{(b)}
   \hfill}
\vspace{0.31\textwidth}    
              
\caption{Solar differential rotation profiles (black lines) from KSO sunspot groups data for solar cycle No. 19 (1954 – 1964), determined by: a) DS method (Table~\ref{table1}, row 1) and b) rLSQ method (Table~\ref{table1}, row 4). The mean values and standard deviations of the sidereal rotation rates within 2-degree bins are represented by the dots and error bars. Box-and-whisker plots show interquartile ranges for bins of unequal heliographic latitude “interval widths”, while keeping the number of sidereal velocities unchanged. The median values are marked with orange lines. Boxplots are scaled to 80\% of the full width for clarity.
        }
\label{Fig5}
\end{figure}

Solar differential rotation profiles calculated from the KSO sunspot groups rotation velocities determined using the DS method (Fig.~\ref{Fig5}, panel (a)) and using the rLSQ method (Fig.~\ref{Fig5}, panel (b)) for solar cycle No. 19 (1954 – 1964) are shown with black lines. Mean values of the sidereal rotation velocities within the $2^{\circ}$ bins of heliographic latitude are shown with black dots, while box-and-whisker plots show interquartile ranges for bins of unequal heliographic latitude ``interval widths'', while keeping the number of sidereal velocities unchanged. In a boxplot, the height of the box shows the interquartile range, e.g. the middle $50\%$ of the data/distribution, while the whiskers (dashed ``error bars'') at the top and bottom show the highest and lowest values. Outliers, e.g. velocities lower than 11 deg/day and higher than 17 deg/day are excluded. Boxplots with the smaller heights indicate less dispersed velocity distributions, while boxplots with higher heights reflect greater dispersion. Also, reaching to the edges of the profile, i.e. to the maximum latitudes at which sunspots are observable (grey boxes)  or to the center of the profile, i.e. the latitudes around the equator (orange and blue box), the boxes become wider. This is not surprising as the nonuniformity of the distribution of sunspots and sunspot groups in different phases of the solar cycle as well as their rarity at higher latitudes are well known \citep{hathaway2015}, making it difficult to measure the solar rotation, especially during the minimum of the activity.


\begin{figure}   
\centerline{\hspace*{0.015\textwidth}
         \includegraphics[width=0.815\textwidth,clip=]{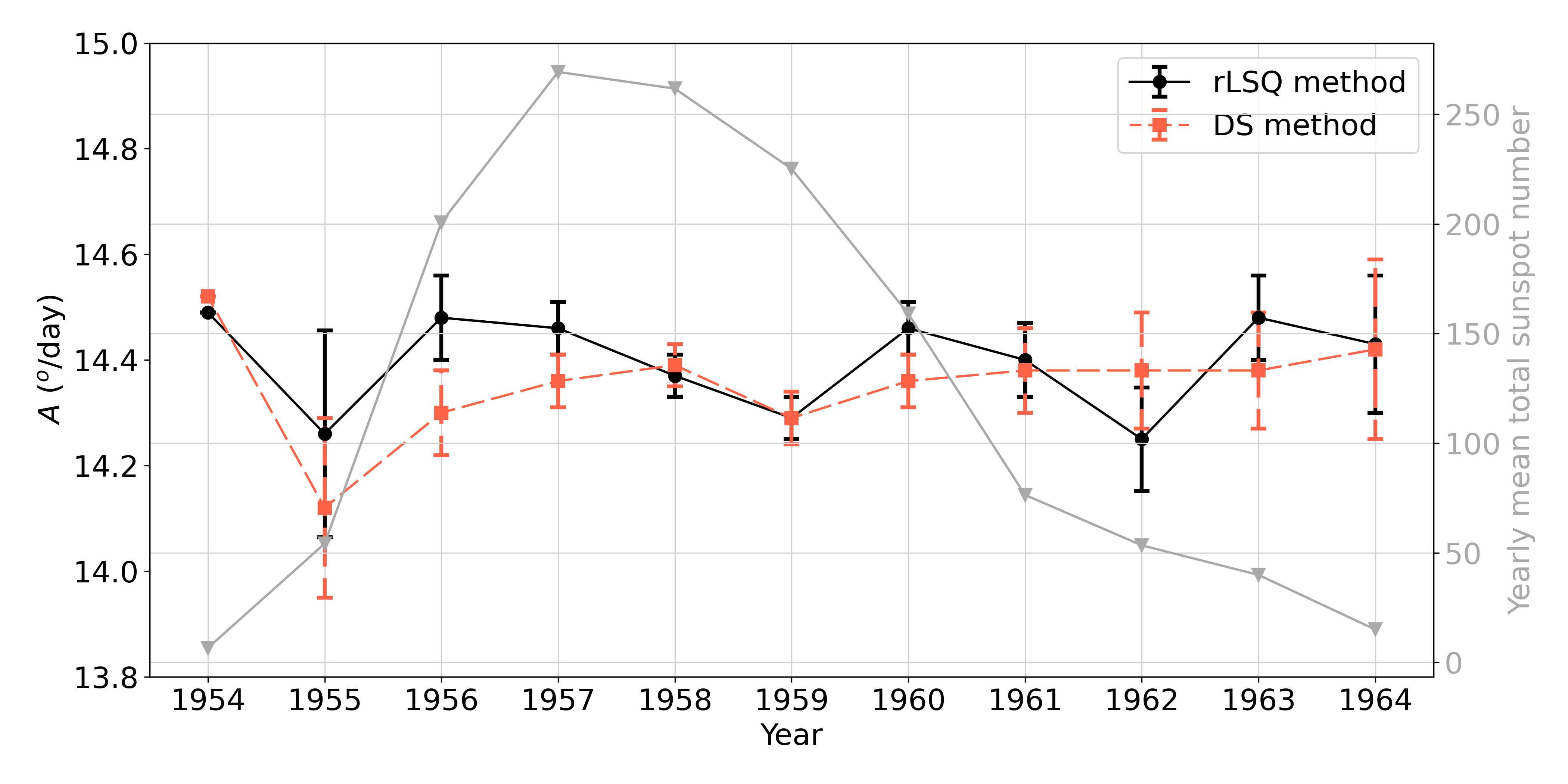}
         \hspace*{-0.03\textwidth}
        }
\vspace{-0.35\textwidth}   
\centerline{\Large \bf     
\hspace{0.0\textwidth} \color{black}{(a)}
   \hfill}
\vspace{0.29\textwidth}    

\centerline{\hspace*{0.02\textwidth}
         \includegraphics[width=0.800\textwidth,clip=]{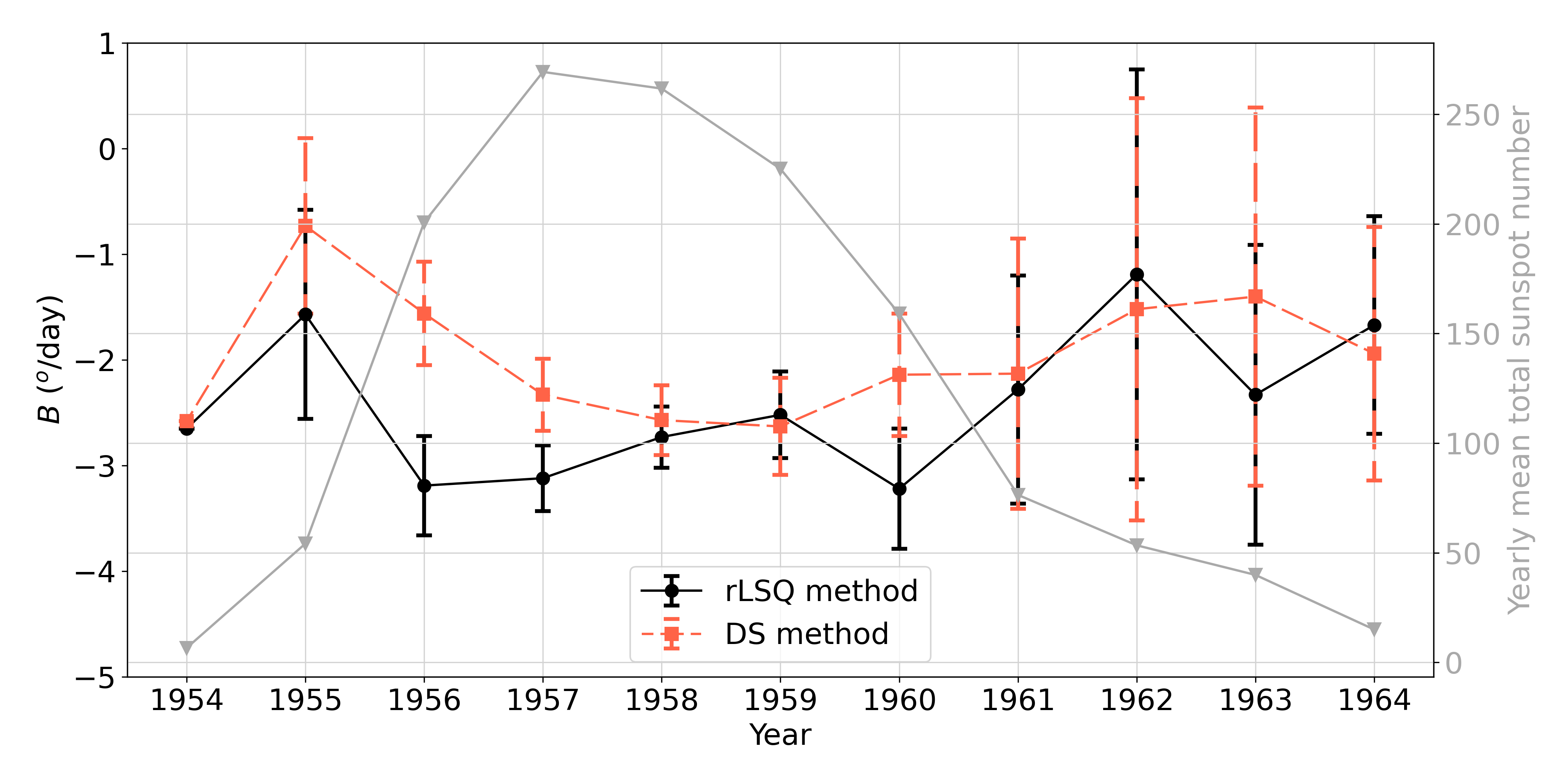}
         \hspace*{-0.03\textwidth}
        }
\vspace{-0.35\textwidth}   
\centerline{\Large \bf     
\hspace{0.0 \textwidth} \color{black}{(b)}
   \hfill}
\vspace{0.29\textwidth}    

\centerline{\hspace*{0.001\textwidth}
         \includegraphics[width=0.830\textwidth,clip=]{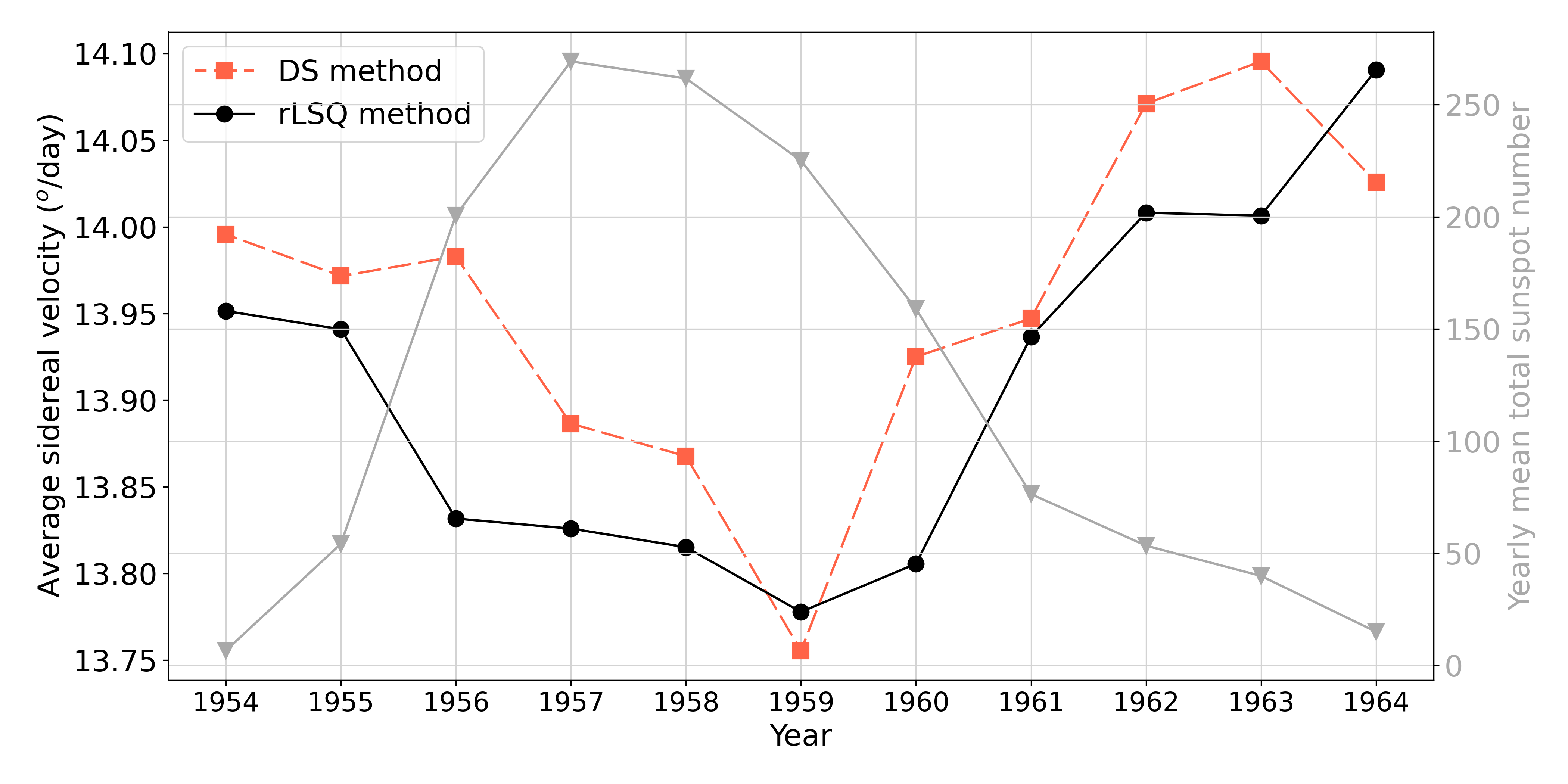}
         \hspace*{-0.03\textwidth}
        }
\vspace{-0.35\textwidth}   
\centerline{\Large \bf     
\hspace{0.0 \textwidth} \color{black}{(c)}
   \hfill}
\vspace{0.29\textwidth}    
              
\caption{Time dependence of the differential rotation parameter $A$ (panel (a)), differential rotation parameter $B$ (panel (b)) and average sidereal velocity (panel (c)) for the whole Sun (N+S) during the solar cycle No. 19 (rLSQ method – full black line with circles, DS method - dashed red line with squares, SILSO yearly mean total sunspot numbers - full grey line with triangles).
        }
\label{Fig6}
\end{figure}

Fig.~\ref{Fig6} shows the differential rotation parameters and average sidereal rotation velocity individually for all the analysed years (1954–1964), marked with black circles (rLSQ method) and red squares (DS method), while the error bars represent the corresponding standard errors. Grey triangles mark the SILSO reclibrated yearly mean total sunspot numbers \citep{clette2014,clette2015,clette2015sac}, i.e. the version 2.0 of the data series provided by \citet{sidc}, Royal Observatory of Belgium, Brussels\footnote{\url{https://www.sidc.be/SILSO/datafiles}}.

Due to the small sample size (15), standard errors are not shown for either method in the year 1954. When analysing differential rotation of sunspot groups for a particular year, observations are limited to a narrow band of heliographic latitudes on the Sun where sunspots appear, depending on the phase of the activity cycle during that year. It directly follows that the range of the independent variable from Eq. (\ref{eq1}), $\sin^2b$, will be limited for a particular year. This reduced variability of the independent variable affects the reliability and necessarily reduces the precision of the estimated parameters\footnote{\url{https://www.statology.org/restriction-of-range/}}. The slope of the fit (parameter $B$) is especially affected by the limited range of the sample, while the precision and reliability of parameter $A$ depends more on the north-south asymmetry of active regions. In order to get an impression of the influence of the limited range on a particular parameter, it is sufficient to compare the magnitudes of the standard errors of parameter $A$ (Fig.~\ref{Fig6}, panel (a)) and parameter $B$ (Fig.~\ref{Fig6}, panel (b)) for a particular year. In addition to the range, the sample size has also an impact on the precision and reliability of the estimated parameters. A larger sample will generally improve the quality of the assessment. Expanding the range of the independent variable and increasing the sample size can be achieved by combining data from all years of the 19th solar cycle.

DS values of rotation parameter $A$ are lower than the rLSQ ones, while DS values of rotation parameter $B$ are less negative than the LSQ ones (the DS method indicates more rigid rotation than the rLSQ one), see Fig.~\ref{Fig6}, panela (a) and (b). Slightly higher values of the DS rotation parameter $A$ are observed during the activity minimum (around 1954 and 1964), and lower during the activity maximum, while for rLSQ values such a solar cycle related variation is not conclusive, see Fig.~\ref{Fig6}, panel (a). Our finding that the equatorial rotation velocity is higher during activity minimum confirms the conclusions from earlier studies \citep{lustig1983, gilmanhow1984, balthasarvazquez1986, brajsa2006, jurdana-sepic2011, li2014, badalyan2017, ruzdjak2017, poljancic2022}.

Panel (c) of Fig.~\ref{Fig6} shows the time dependence of the average sidereal velocities, calculated using the equation:
   \begin{equation}
    \label{eq3}
      \bar{\omega_i} = \frac{1}{48^{\circ}-(-48^{\circ})}\int \limits_{-48^{\circ}}^{48^{\circ}}\left(A_i+B_i \sin^2b\right) \mathrm{d}b
   \end{equation}
where i indicates corresponding year, and the average velocity is determined within the range $\pm 48$ deg in heliographic latitude. These limits are chosen according to the extent of the measurements visible in Fig.~\ref{Fig5}.


It is known that sunspot groups evolve both during their lifetime and during the solar cycle, moving from simpler to more complex structures. This evolution is closely linked to an increase in the area and extent of their magnetic activity \citep{chen2011,watsonfm2011}. Sunspot groups also exhibit differences in rotational velocities depending on their size and complexity. Numerous studies have shown that smaller and generally simpler sunspot groups tend to rotate faster, while larger, more complex groups rotate more slowly \citep{howard1984,ruzdjak2004,ruzdjak2005}. Since smaller, fast-rotating groups are more prevalent near solar minima, the average sidereal rotation rate of sunspot groups tends to be higher during these periods. Conversely, during solar maxima, when larger and slower groups dominate, the mean rotation rate decreases. A tendency of this kind is noticeable in Fig.~\ref{Fig6} (panel (c)), where both the DS and rLSQ cases show a decrease in average sidereal rotational velocity during the ascending phase of the solar cycle (1954–1959) and an increase during the descending phase (1959–1964).

\subsection{Comparison with other data sets for cycle No. 19}\label{S-Comparison} 

\begin{figure}    
\centerline{\includegraphics[width=1.05\textwidth,clip=]{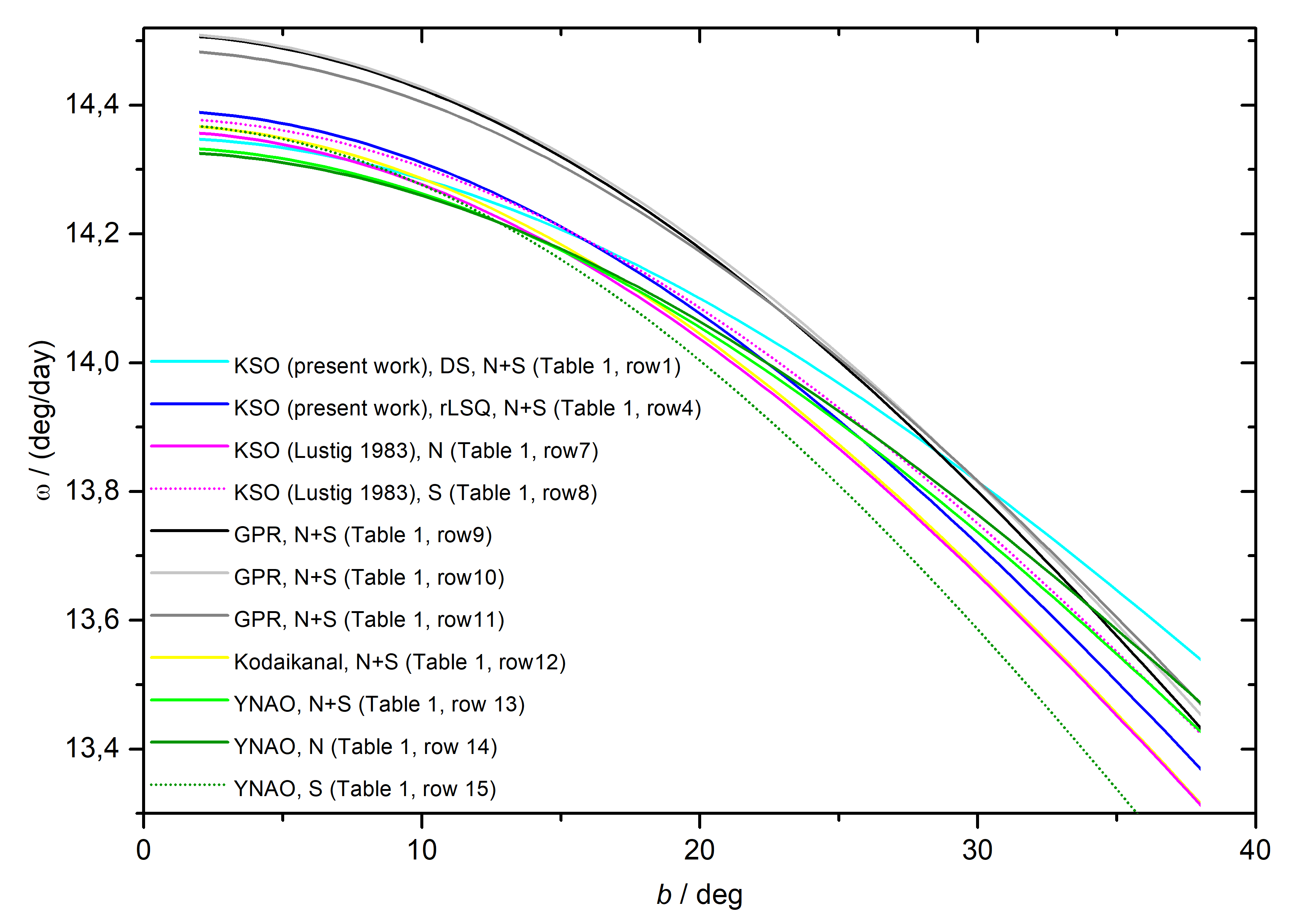}}
\small
        \caption{Differential rotation profiles for solar cycle No. 19, as determined by various authors using multiple data sets (GPR, KoSO, YNAO, KSO). For numerical values of differential rotation parameters refer to corresponding rows in Table~\ref{table1}. The heliographic latitude is represented by $b$, and the sidereal rotation velocity is represented by $\omega$.
                }
\label{Fig7}
\end{figure}


\begin{table*}[ht]
\caption{Comparison of the differential rotation parameters $A$ determined from KSO DS/rLSQ data in present paper (Table~\ref{table1}, rows 1-6) with those determined by different authors for the solar cycle No. 19 (Table~\ref{table1}, rows 7-15). Inequalities arise from Eq. (\ref{eq2}). Green, orange and yellow - compared $A$ values statistically coincide within one, two or three common $\sigma$, red - compared $A$ values are statistically significantly different (more than three common $\sigma$ apart). N+S denotes both hemispheres combined, N represents the northern hemisphere and S the southern hemisphere. Superscripts indicate the row in Table~\ref{table1}.}
\label{table2}
\centering
\begin{tabular}{l>{\centering\arraybackslash}p{1.1cm}>{\centering\arraybackslash}p{1.1cm}>{\centering\arraybackslash}p{1.1cm}>{\centering\arraybackslash}p{1.1cm}>{\centering\arraybackslash}p{1.1cm}>{\centering\arraybackslash}p{1.1cm}}
\hline
Data set, & KSO DS, & KSO DS, & KSO DS, & KSO rLSQ,  & KSO rLSQ,  & KSO rLSQ,  \\
Hemisphere & N+S$^1$  & N$^2$ & S$^3$ & N+S$^4$ & N$^5$ & S$^6$ \\
\hline
KSO, N$^7$ &  &\textcolor{green}{$<\!1\sigma$}  & & & \textcolor{green}{$<\!1\sigma$} & \\
KSO, S$^8$  & & &\textcolor{green}{$<\!1\sigma$} & & & \textcolor{orange}{$<\!2\sigma$} \\
GPR, N+S$^9$  &\textcolor{red}{$>\!3\sigma$} & & &\textcolor{red}{$>\!3\sigma$} & & \\
GPR, N+S$^{10}$  & \textcolor{red}{$>\!3\sigma$} & & & \textcolor{red}{$>\!3\sigma$}& & \\
GPR, N+S$^{11}$ & \textcolor{red}{$>\!3\sigma$} & & & \textcolor{red}{$>\!3\sigma$}& & \\
KoSO, N+S$^{12}$ & \textcolor{orange}{$<\!2\sigma$} & & & \textcolor{green}{$<\!1\sigma$} & & \\ 
YNAO, N+S$^{13}$ & \textcolor{green}{$<\!1\sigma$} & & & \textcolor{orange}{$<\!2\sigma$} & & \\
YNAO, N$^{14}$  & & \textcolor{green}{$<\!1\sigma$} & & & \textcolor{green}{$<\!1\sigma$}& \\
YNAO, S$^{15}$ & & & \textcolor{green}{$<\!1\sigma$}& & & \textcolor{orange}{$<\!2\sigma$} \\
\hline
\end{tabular}
\end{table*}


\begin{table*}[ht]
\caption{The same as in Table~\ref{table2} but for differential rotation parameter $B$.}
\label{table3}
\centering
\begin{tabular}{l>{\centering\arraybackslash}p{1.1cm}>{\centering\arraybackslash}p{1.1cm}>{\centering\arraybackslash}p{1.1cm}>{\centering\arraybackslash}p{1.1cm}>{\centering\arraybackslash}p{1.1cm}>{\centering\arraybackslash}p{1.1cm}}
\hline
Data set, & KSO DS, & KSO DS, & KSO DS  & KSO rLSQ  & KSO rLSQ  & KSO rLSQ  \\
Hemisphere & N+S$^1$  & N$^2$ & S$^3$ & N+S$^4$ & N$^5$ & S$^6$ \\
\hline
KSO N$^7$  &  &\textcolor{orange}{$<\!2\sigma$}  & & & \textcolor{green}{$<\!1\sigma$} & \\
KSO S$^8$ & & &\textcolor{green}{$<\!1\sigma$} & & &  \textcolor{green}{$<\!1\sigma$} \\
GPR N+S$^9$ & \textcolor{goldenyellow}{$<\!3\sigma$} & & &\textcolor{green}{$<\!1\sigma$} & & \\
GPR N+S$^{10}$ & \textcolor{goldenyellow}{$<\!3\sigma$} & & & \textcolor{green}{$<\!1\sigma$}& & \\
GPR N+S$^{11}$ & \textcolor{goldenyellow}{$<\!3\sigma$} & & & \textcolor{green}{$<\!1\sigma$}& & \\
KoSO N+S$^{12}$ & \textcolor{green}{$<\!1\sigma$} & & & \textcolor{green}{$<\!1\sigma$} & & \\ 
YNAO N+S$^{13}$ & \textcolor{green}{$<\!1\sigma$} & & & \textcolor{green}{$<\!1\sigma$} & & \\
YNAO N$^{14}$ & & \textcolor{green}{$<\!1\sigma$} & & & \textcolor{green}{$<\!1\sigma$}& \\
YNAO S$^{15}$ & & & \textcolor{green}{$<\!1\sigma$}& & & \textcolor{green}{$<\!1\sigma$}  \\
\hline
\end{tabular}
\end{table*}

The differential rotation parameters collected from different sources (KSO, GPR, KoSO, YNAO) for solar cycle No. 19, using only sunspots and sunspot groups as tracers, are listed in Table~\ref{table1} (rows 7-15).  Fig.~\ref{Fig7} shows the corresponding differential rotation profiles. Comparison of the differential rotation parameters $A$ determined by KSO DS/rLSQ data in present paper (Table~\ref{table1}, rows 1-6) and the ones determined by different authors for the solar cycle No. 19 (Table~\ref{table1}, rows 7-15) is available in Table~\ref{table2}. The same, but for differential rotation parameter $B$ is available in Table~\ref{table3}.

Inequalities in the Tables \ref{table2} and \ref{table3} arise from Eq. (\ref{eq2}). Inequalities compare the change in the two differential rotation parameters and the sum of the corresponding standard errors of the parameters. Red indicates that the change between the two parameters is larger than the sum of the threefold standard errors of the compared parameters, i.e. that compared parameters are statistically significantly different. Yellow/orange/green inequalities indicate that the change between the two parameters is smaller than the sum of the threefold/twofold/onefold standard errors of the compared parameters, i.e. that compared values are statistically identical (statistically significant coincidence is present).

In Table~\ref{table2} only the comparison of KSO DS/rLSQ results from the present paper and GPR results from three different papers yield statistically significant difference (red inequalities). The rest of the table is in green, orange and yellow, which means that the results for equatorial rotation velocity (parameter $A$) are statistically identical (significantly coincide) for the pairs of KSO DS/rLSQ data from the present paper and all other data sets:  KSO \citep{lustig1983}, KoSO \citep{jha2021} and YNAO \citep{luo2021}. This holds no matter whether the entire Sun or individual hemispheres are compared. In Fig.~\ref{Fig7}, it is clearly visible that the profiles of these data sets are densely interwoven at lower heliographic latitudes (near the equator). Also, one can notice that there is a gap between the stripe of the three different studies based on the GPR data and the stripe of the rest (KSO, KoSO, YNAO).

Lower values for $A$ from KSO compared to GPR were previously observed in \citet{balthasar1984} and \citet{poljancic2017}. \citet{balthasar1984} identified two main differences: 1) the solar equatorial rotation velocities from KSO data are generally smaller than those from GPR data, and 2) KSO data show an increase in solar equatorial rotation velocity of about 0.1 deg/day per decade, whereas GPR data show no long-term trend. They attributed these discrepancies entirely to instrumental factors and proposed numerical corrections to the KSO results from \citet{lustig1983}. However, concerning the findings for cycle No. 19 in the present paper, the GPR data show the greatest inconsistency compared to the other three data sources. It is implausible that three different data sets - KSO, YNAO, and KoSO - could share the same systematic errors as those reported for KSO in the work of \citet{balthasar1984}. Therefore, the causes of these differences need to be further investigated by carefully analysing the methods used to determine the sunspot group positions of several different data sets over longer periods of time. The discrepancy in the rotation rate between the GPR data and other sources may be explained by differences in the procedures used to determine the positions of sunspot group centers. The center can be defined as the geometrical center, an area-weighted center, or an intensity-weighted center. These approaches can yield different results, as the determined position often depends on the sunspot group’s morphology and the specific method applied. Therefore, we provide a center-determination procedures used in the YNAO, KoSO, and GPR datasets. A more detailed description of the method applied on KSO data is provided in Section~\ref{S-Methods}. 

To determine the central position of sunspot groups in the YNAO dataset \citep[Section~2.2]{luo2021}, a semi-automatic program was used. The Hough transform, a classical computer vision technique for detecting the general low-parametric objects such as circles, is used to get the circle on the hand-drawing sunspot records. The coordinates of the centre of gravity $(B_p, L_p)$ of the sunspot group are calculated and marked in the figure. Taking into account the time difference between the hand-drawing image and the satellite image at the closest time, the position of the hand-drawing sunspot group is corrected by the classical differential rotation formula.

For the determination of the positions of the sunspots on KoSO calibrated white light images \citep[Section~4]{mandal2017}, a modified version of the ``sunspot tracking and recognition algorithm'' (STARA) has been used. They modified the code to get the position information, longitude and latitude, from the ``center of gravity'' method of the detected sunspots. 

The exact method of measuring the GPR positions and areas of sunspots is described in a series of publications that constitute the Greenwich Photoheliographic Results 1873-1976, available via UK solar system data centre\footnote{\url{https://www.ukssdc.ac.uk/wdcc1/RGOPHR/}}, as well as within the \citet{willis1996} and \citet{willis2013}. For large or complex sunspot groups the position of the ``center of gravity'' has been calculated in such a way that the longitude and latitude of each chief components of the sunspot group were measured individually. Then, the ``center of gravity'' has been calculated by multiplying the longitude and latitude of each separately measured component of the group by its area, and dividing the sum of the products by the sum of the areas. In other cases (simple unipolar groups) the position of the center of the group was estimated by the measurer using the micrometer.

Since all observatories use the same procedure for determining positions—the ``center of gravity'' or area-weighted method—it appears that the method of center determination does not account for the observed difference in the value of parameter $A$ shown in Fig.~\ref{Fig7}. Therefore, as a next step, a more detailed analysis should be conducted to investigate the cause of the statistically significant difference in equatorial rotation velocities observed by GPR and the other datasets (YNAO, KoSO, and KSO). The plan also includes analyzing the behavior across other available solar cycles (Nos. 20–23) to investigate whether a long-term increasing trend in the values of parameter $A$ exists, as was proposed for KSO by \citet{balthasar1984}.

The comparison for parameter $B$ (Table~\ref{table3}) yield statistically identical results in all cases. Fig.~\ref{Fig7} clearly shows that the profiles of all data sets merge at larger heliographic latitudes, i.e. the degree of rotational non-uniformity is approximately the same.

To assess the validity of our visual method for determining sunspot group positions, we highlight Fig.~\ref{Fig2}, which shows several sunspot groups with both our visually determined positions (green) and the corresponding GPR positions (blue), derived from nearly simultaneous observations (November 26, 1958 – only 8 minutes apart). It is evident that, despite using a visual method, our positions agree quite well with those from GPR.

\begin{figure}   
\centerline{\hspace*{0.015\textwidth}
         \includegraphics[width=.915\textwidth,clip=]{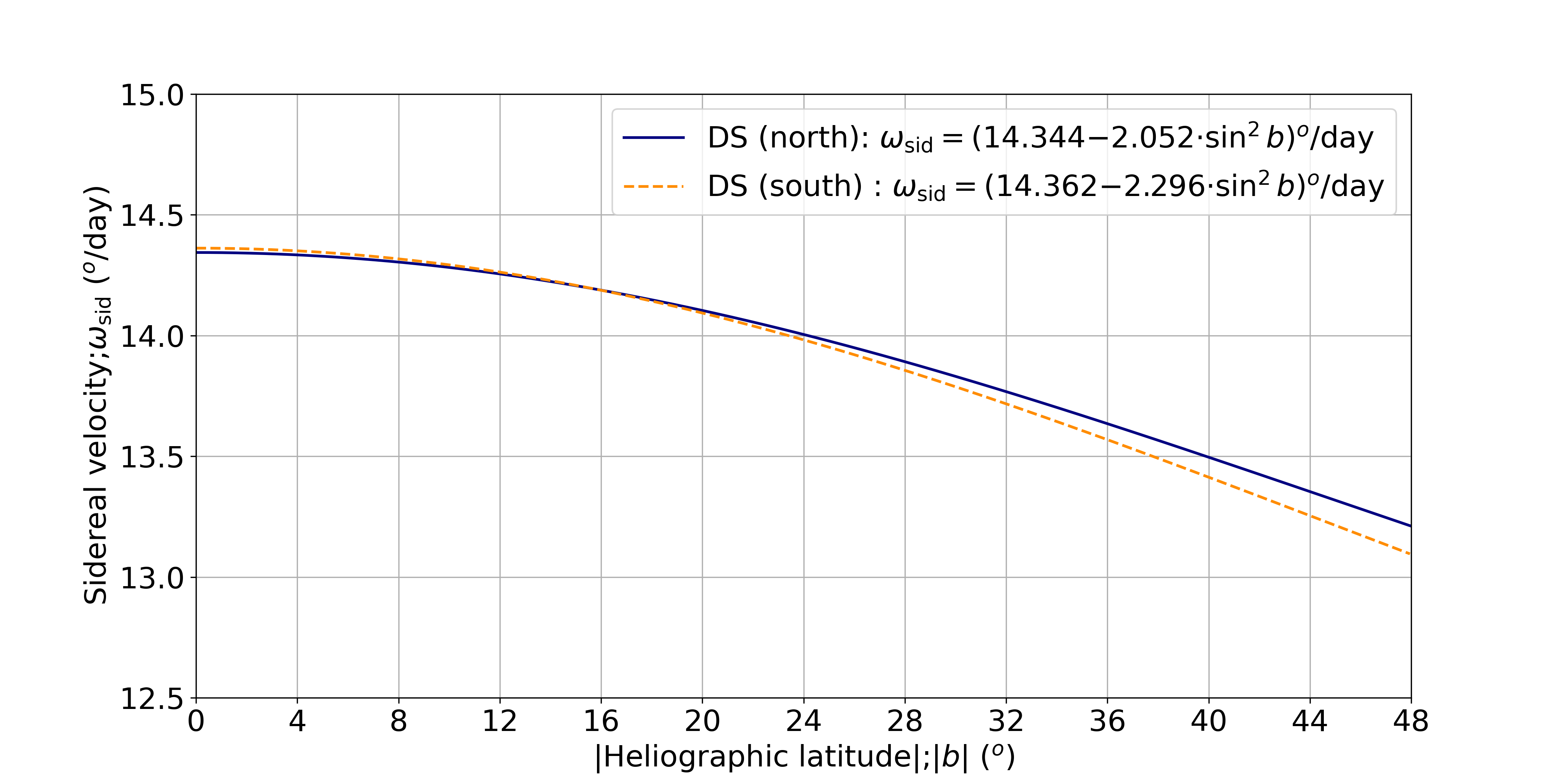}
         \hspace*{-0.03\textwidth}
        }
\vspace{-0.37\textwidth}   
\centerline{\Large \bf     
\hspace{0.0\textwidth} \color{black}{(a)}
   \hfill}
\vspace{0.32\textwidth}    
\centerline{\hspace*{0.015\textwidth}
         \includegraphics[width=0.915\textwidth,clip=]{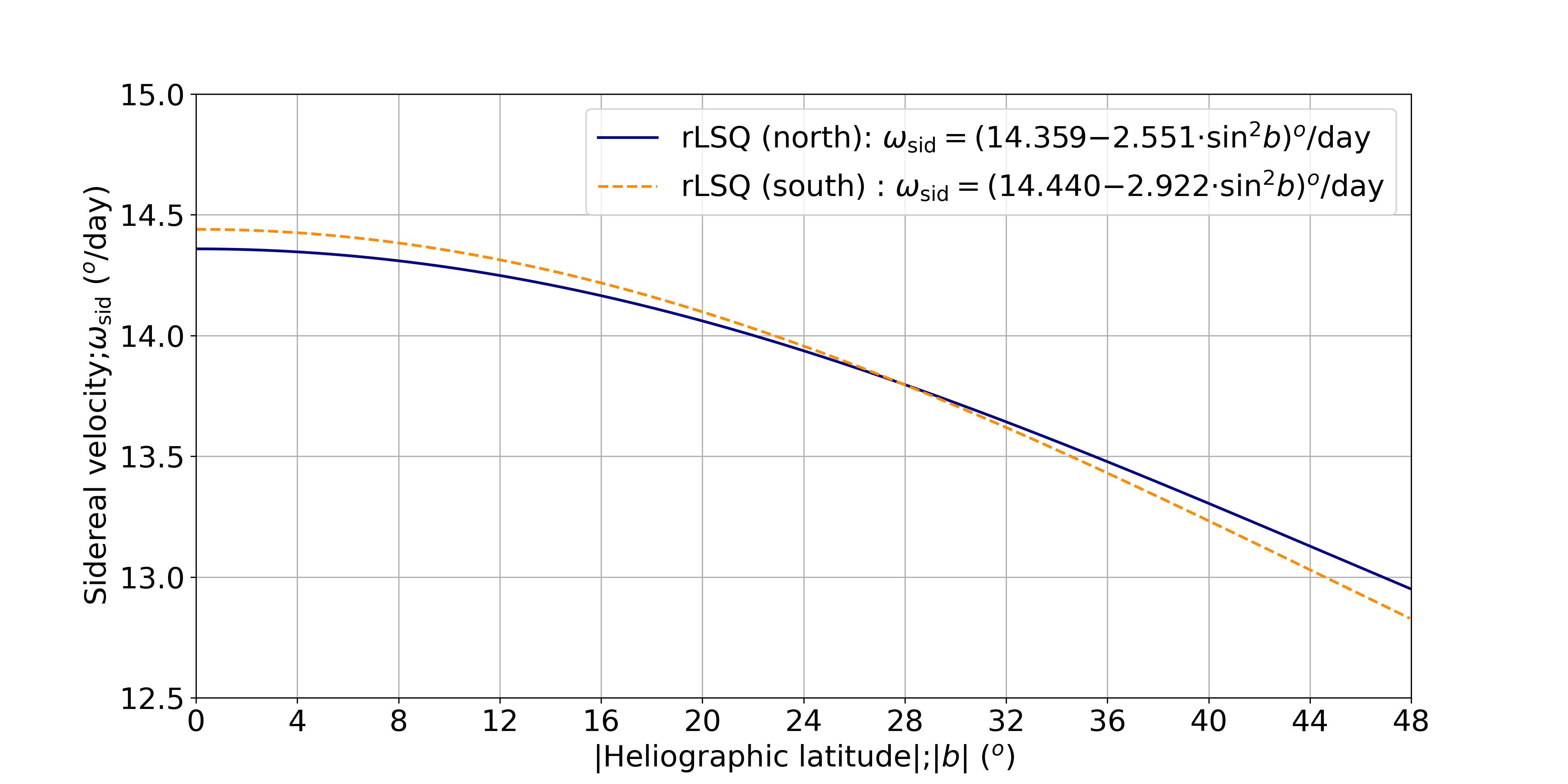}
         \hspace*{-0.03\textwidth}
        }
\vspace{-0.35\textwidth}   
\centerline{\Large \bf     
\hspace{0.0 \textwidth} \color{black}{(b)}
   \hfill}
\vspace{0.31\textwidth}    
              
\caption{Solar rotation profiles derived separately for the Northern (blue) and Southern (orange) hemispheres for solar cycle No. 19 (1954–1964): a) DS method (Table~\ref{table1}, rows 2 and 3) and b) rLSQ method (Table~\ref{table1}, rows 5 and 6). The heliographic latitude is represented by $b$, and the sidereal rotation velocity is represented by $\omega$.}
\label{Fig8}
\end{figure}

\subsection{North–south asymmetry of the solar rotation for cycle No. 19}\label{S-NSasymmetry} 

The differential rotation parameters derived from the northern and southern KSO hemispheric data for solar cycle No. 19 (1954–1964) are given in Table~\ref{table1} (DS method – rows 2 and 3; rLSQ method – rows 5 and 6). The differential rotation profiles are shown in Fig.~\ref{Fig8}. 

When the corresponding values of the northern and southern differential rotation parameters (Table~\ref{table1}: row 2 vs. row 3, row 5 vs. row 6) are compared, according to Eq. (\ref{eq2}), the derived change between the two parameters is smaller than the sum of the twofold/onefold standard errors of the compared parameters. This means that the compared values significantly coincide. Nevertheless, it can be seen in Fig.~\ref{Fig8} that the equatorial rotation velocity is slightly higher in the southern hemisphere (the orange lines lie slightly above the blue lines at low heliographic latitudes) and that the southern hemisphere rotates slightly more differentiated (a slightly more negative value of parameter $B$ is present). This applies to both methods used, DS and rLSQ.

\begin{figure}   
\centerline{\hspace*{0.015\textwidth}
         \includegraphics[width=.915\textwidth,clip=]{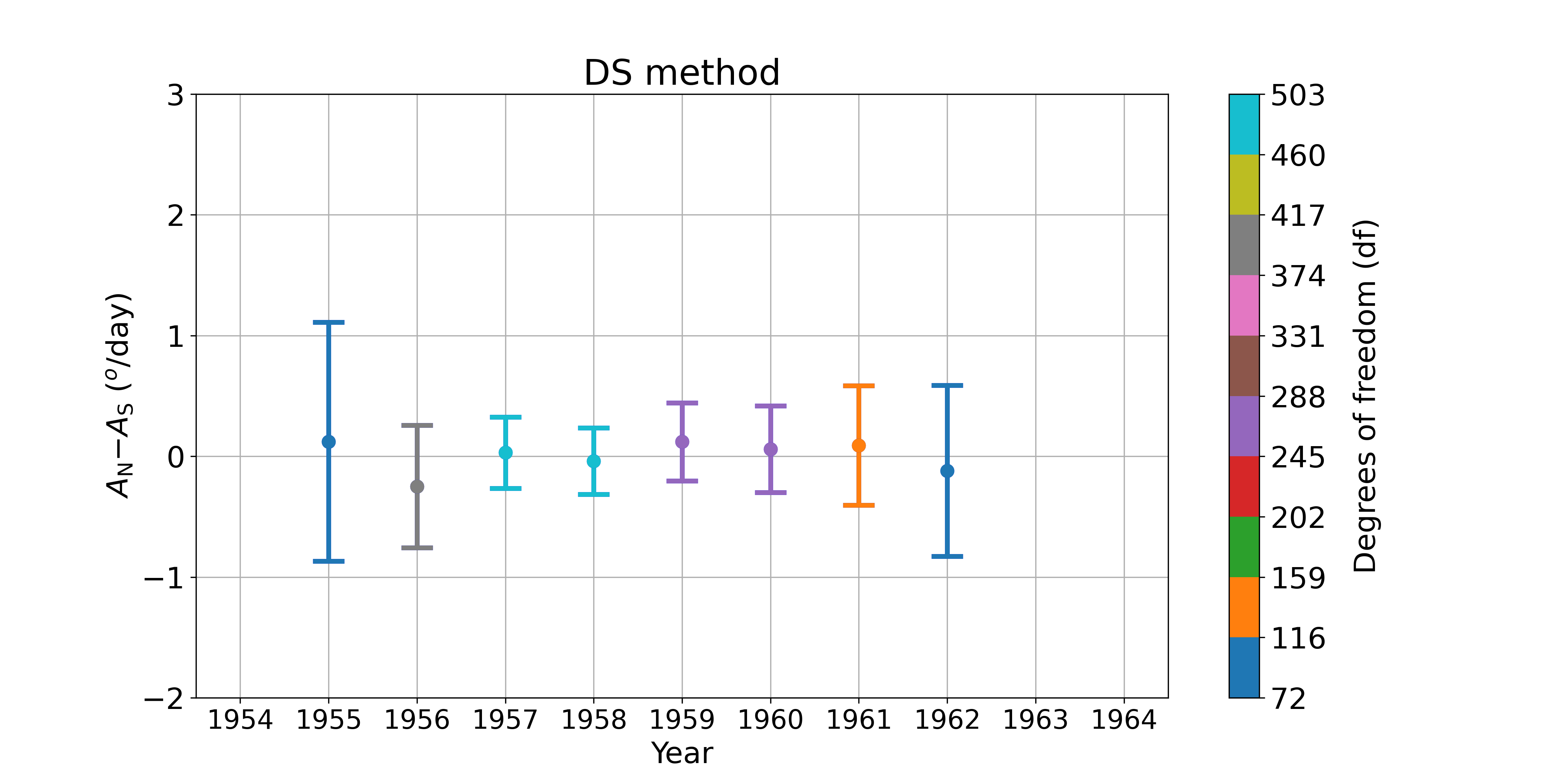}
         \hspace*{-0.03\textwidth}
        }
\vspace{-0.37\textwidth}   
\centerline{\Large \bf     
\hspace{0.0\textwidth} \color{black}{(a)}
   \hfill}
\vspace{0.32\textwidth}    
\centerline{\hspace*{0.015\textwidth}
         \includegraphics[width=0.915\textwidth,clip=]{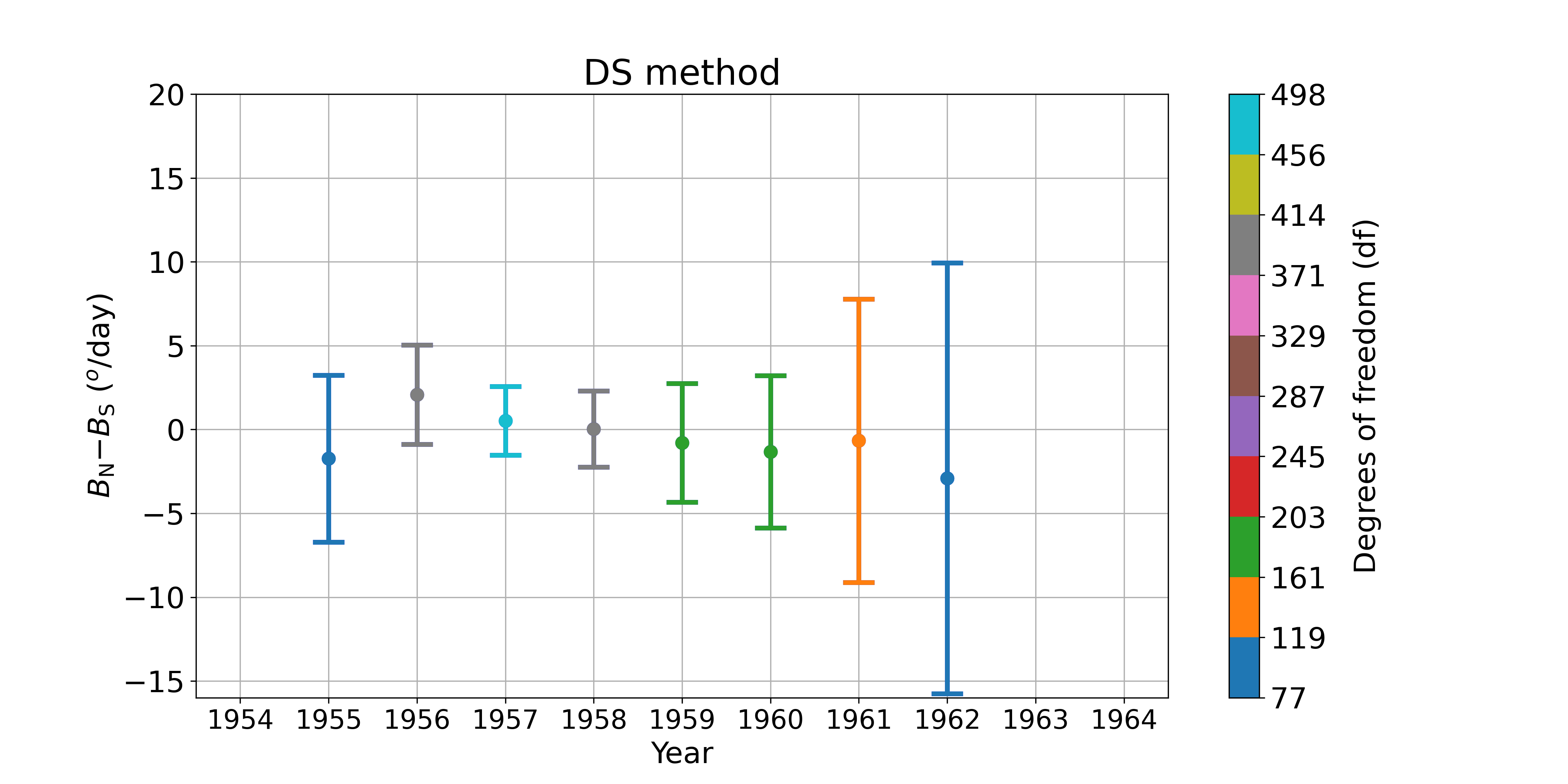}
         \hspace*{-0.03\textwidth}
        }
\vspace{-0.35\textwidth}   
\centerline{\Large \bf     
\hspace{0.0 \textwidth} \color{black}{(b)}
   \hfill}
\vspace{0.31\textwidth}    

\caption{Confidence intervals of the difference between values of differential rotation parameters ($A$, $B$) of the northern and southern hemispheres for a particular year (DS method). Degrees of freedom (df) represents the number of samples.}\label{Fig9}
\end{figure}


\begin{figure}   
\centerline{\hspace*{0.015\textwidth}
         \includegraphics[width=.915\textwidth,clip=]{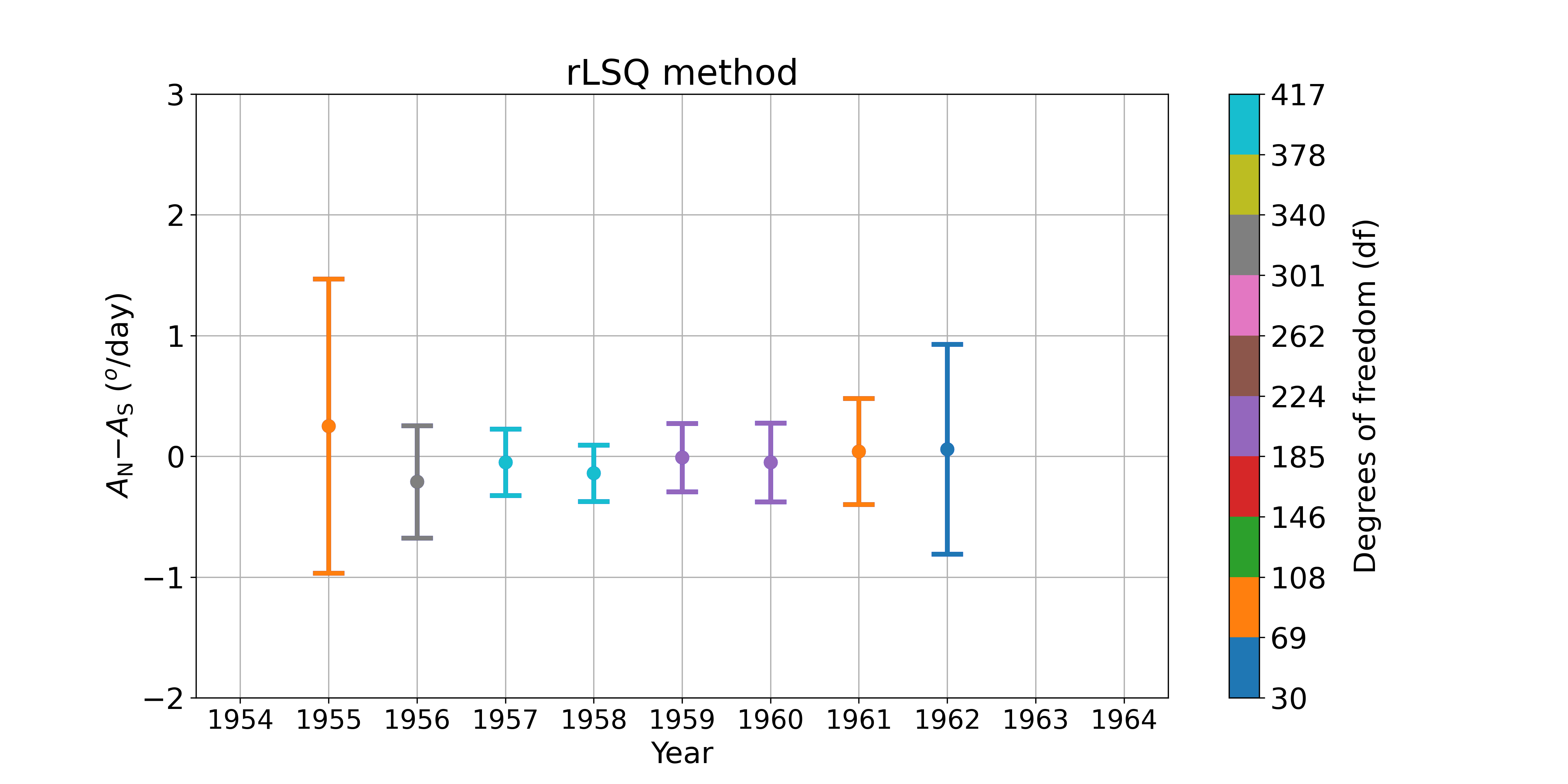}
         \hspace*{-0.03\textwidth}
        }
\vspace{-0.37\textwidth}   
\centerline{\Large \bf     
\hspace{0.0\textwidth} \color{black}{(a)}
   \hfill}
\vspace{0.32\textwidth}    
\centerline{\hspace*{0.015\textwidth}
         \includegraphics[width=0.915\textwidth,clip=]{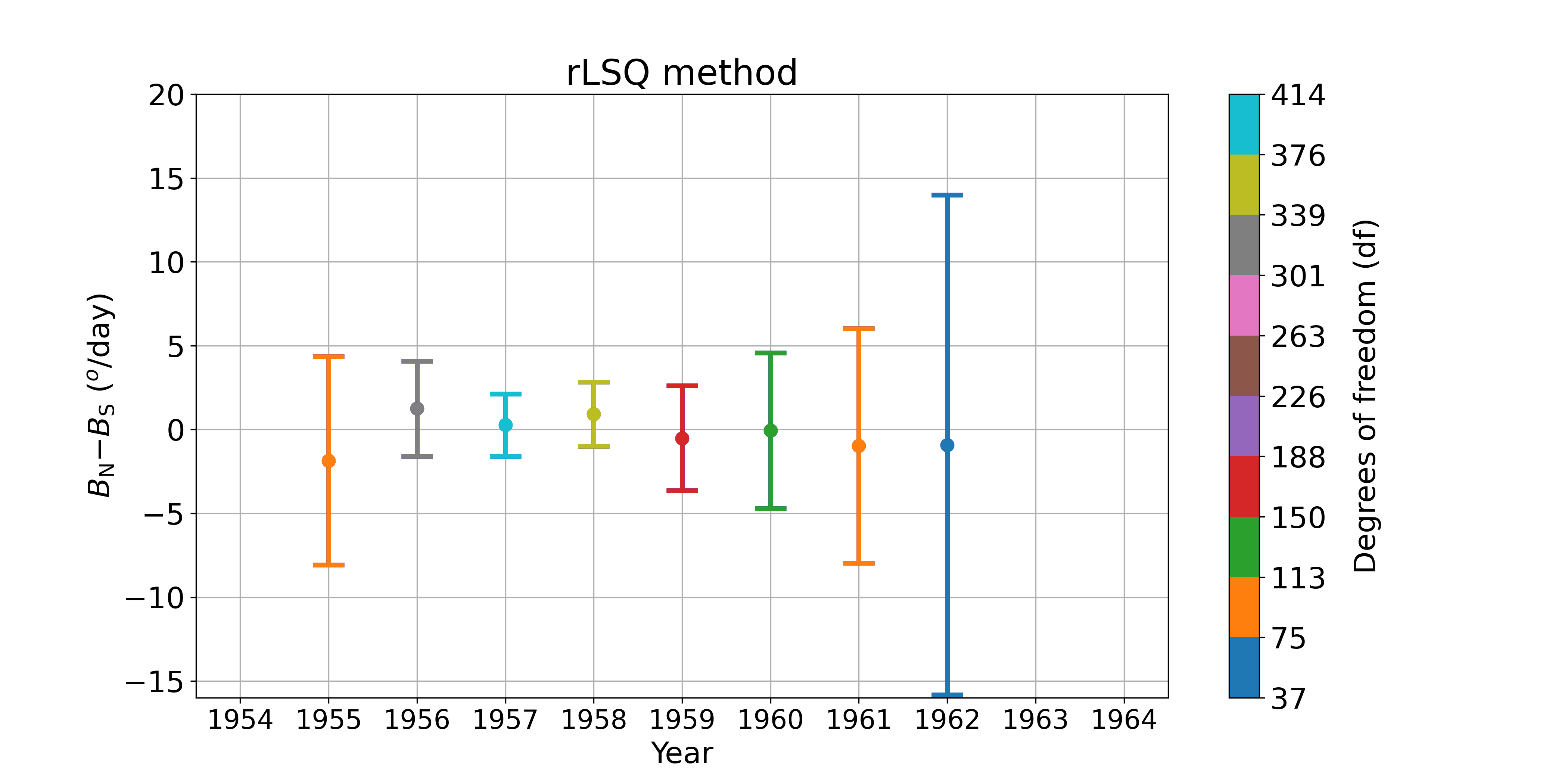}
         \hspace*{-0.03\textwidth}
        }
\vspace{-0.35\textwidth}   
\centerline{\Large \bf     
\hspace{0.0 \textwidth} \color{black}{(b)}
   \hfill}
\vspace{0.31\textwidth}    

\caption{The same as in Fig.~\ref{Fig9} but for rLSQ method.}\label{Fig10}
\end{figure}

The same is obtained when the results for the individual years are analysed. Fig.~\ref{Fig9} (DS method) and Fig.~\ref{Fig10} (rLSQ method) show the $99.7\%$ confidence intervals of the difference between values of differential rotation parameters ($A, B$) of the northern (label N) and southern (label S) hemispheres for a particular year. Confidence intervals (central intervals with $\alpha = 0.003$) for two independent samples, the parameters of the northern and southern solar hemispheres, were constructed from the Student's $t$-distribution with df degrees of freedom (color bars in Figs.~\ref{Fig9}~and~\ref{Fig10}), as follows
   \begin{equation}
    \label{eq4}
      (\beta_N-\beta_S) = (\overline{\beta}_N-\overline{\beta}_S) \pm t_{\frac{\alpha}{2},\mathrm{df}}\cdot \sqrt{SE_N^2+SE_S^2}
   \end{equation}
where $\beta$ is the general notation for the regression parameter $A$ or $B$, $\overline{\beta}_N-\overline{\beta}_S$ is the difference between values of differential rotation parameters and $SE_N$ ($SE_S$) is the standard error associated with the differential rotation parameter $\overline{\beta}_N$ ($\overline{\beta}_S$) of the northern (southern) hemisphere. Due to the small sample size, the corresponding points and confidence intervals are not shown for 1954, 1963, and 1964 in either method. It is evident that all confidence intervals from Figs.~\ref{Fig9}~and~\ref{Fig10} include zero which indicates that there is no statistically significant north–south asymmetry of the solar rotation during the individual years.

When the same statistical test is performed for the whole solar cycle No. 19, for both parameters and both methods (Table~\ref{table4}), the absolute value of the $t$-value calculated from the sample is always less than the critical value with a significance level of $0.003$. Therefore, the difference is not statistically significant and the null hypothesis (symmetry in the rotation of the Sun around the equator during the cycle No. 19) cannot be rejected. 

\begin{table*}[ht]
\caption{Statistical objects for testing north–south asymmetry of the solar rotation for cycle No. 19.}
\label{table4}
\centering
\begin{tabular}{c>{\centering\arraybackslash}p{1.4cm}>{\centering\arraybackslash}p{2.0cm}>{\centering\arraybackslash}p{1.4cm}>{\centering\arraybackslash}p{1.4cm}c}
\hline
Method & Difference& Confidence interval (°/day) & Absolute $t$-value; $t_\mathrm{exp}$ & Critical $t$-value; $t_0$ & $|t_\mathrm{exp}|<t_0$\\
\hline
rLSQ   & $A_\mathrm{N}-A_\mathrm{S}$ & $[-0.195,0.033]$ & $2.110$ & $2.971$  & True \\
rLSQ   & $B_\mathrm{N}-B_\mathrm{S}$ & $[-0.529,1.271]$ & $1.225$ & $2.972$  & True\\
DS   & $A_\mathrm{N}-A_\mathrm{S}$ & $[-0.145,0.109]$ & $0.422$     & $2.971$ & True\\
DS   & $B_\mathrm{N}-B_\mathrm{S}$ & $[-0.745,1.233]$ & $0.733$   & $2.971$  & True\\
\hline
\end{tabular}
\end{table*}

\begin{figure}    
\centerline{\includegraphics[width=1.05\textwidth,clip=]{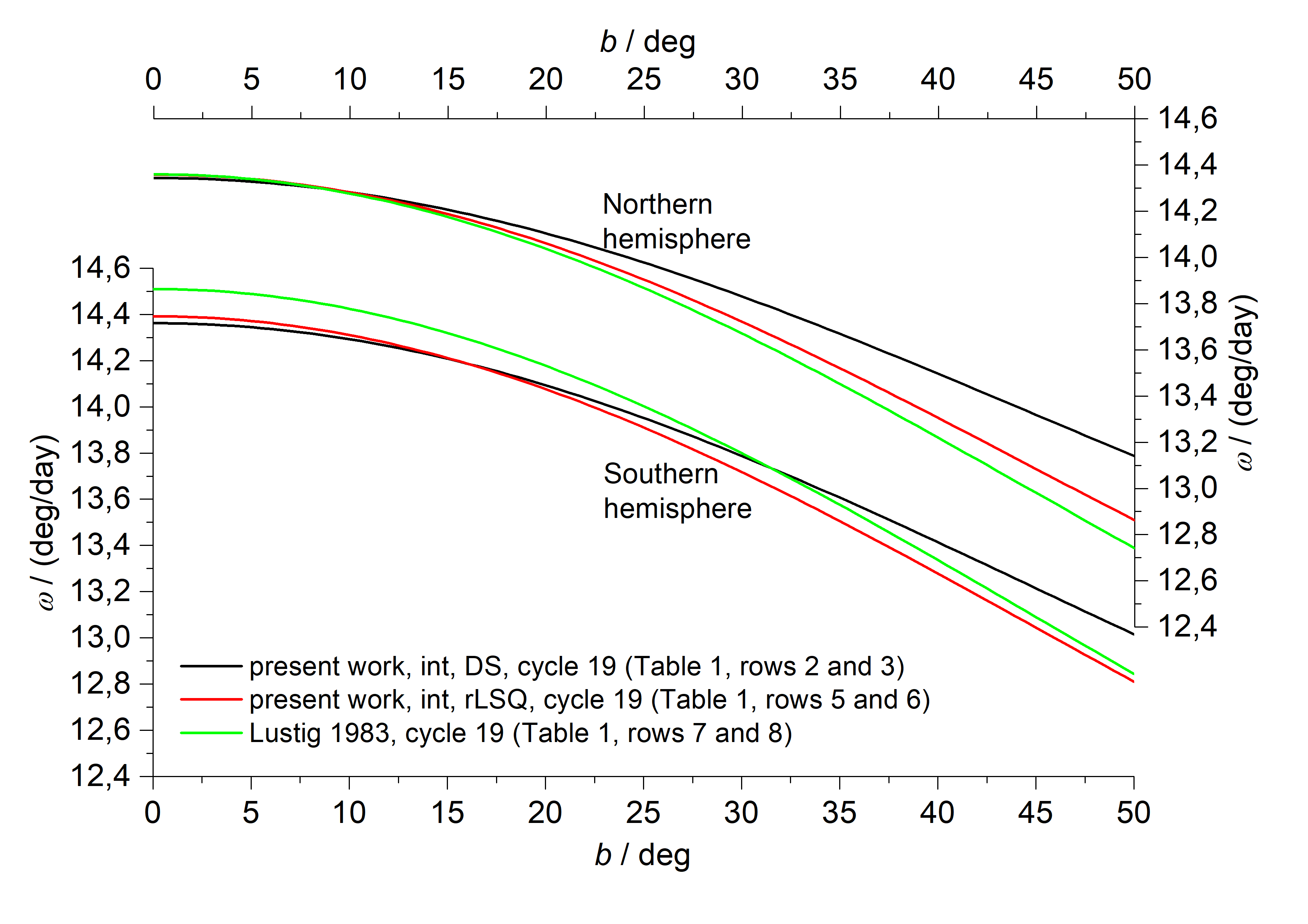}}
\small
        \caption{The hemispheric differential rotation profiles calculated for solar cycle No. 19 (1954–1964) in present paper and \citet{lustig1983}. Exact values of parameters $A$ and $B$ for the northern and southern hemispheres are available in corresponding rows of Table~\ref{table1}. The heliographic latitude is represented by $b$, and the sidereal rotation velocity is represented by $\omega$.
                }
\label{Fig11}
\end{figure}

When the hemispheric differential rotation profiles for solar cycle No. 19 (1954–1964) from \citet{lustig1983} are compared, using rows 7 and 8 in Table~\ref{table1} and Eq. (\ref{eq2}), statistically insignificant differences are obtained for the equatorial velocities (the parameter $A$), as well as for the gradients of the differential rotation (parameter $B$), between the northern and southern hemispheres. The same result is obtained when the values from the YNAO hemispheric data are compared (Table~\ref{table1}, rows 14 and 15). Therefore, the conclusion is the same as for the results from the present work: the compared values for the two hemispheres significantly coincide and indeed the symmetry in the rotation of the Sun around the equator during cycle No. 19 is confirmed. Nevertheless, the KSO data of \citet{lustig1983} and the YNAO data also show slightly higher values for $A$ in the southern hemisphere compared to the northern hemisphere, as was also found for the KSO results in the present paper. The corresponding rotation profiles from \citet{lustig1983} and present paper are shown in Fig.~\ref{Fig11}.

Upon completion of the KSO catalog of positions and rotation velocities, a detailed analysis of the north-south asymmetry in the solar rotation profile of the KSO data and its comparison with the north-south asymmetry of the activity of the KSO data \citep{temmer2006} and the WDC SIDC (ROB) data is planned. We hope this will offer a practical explanation for the intriguing results of solar cycle No. 19, including the pronounced north-south asymmetry in activity \citep{temmer2006,veronig2021} and the symmetry in solar rotation revealed in this study.

\section{Conclusions}\label{S-Conclusions} 

The positional data derived from the sunspot drawings are a valuable resource for analysing the long-term variability of solar rotation, the north–south asymmetry of rotation and activity and the relationship between solar rotation and activity. Such analyses help to interpret and understand the solar dynamo. With the discontinuation of DPD and SOON/USAF/NOAA, the well-known datasets that continued the legacy of the GPR, it is crucial to fill the newly created gaps and develop another user-friendly datasets.

The KSO dataset, which agrees well with both the DPD and the GPR on a half-century time scale (1964-2016, as shown in previous studies), is a strong candidate for this purpose. Accordingly, we present results on sunspot group positions, photospheric differential rotation parameters, their temporal variations, comparison with the results from other available datasets and north–south rotational asymmetry for solar cycle No. 19 (1954–1964). These results were derived from tracing sunspot groups on KSO sunspot drawings and extend the differential rotation analysis of KSO data from 1964 to 2016 \citep{poljancic2017}. Since the first year in which sunspot drawings appear at KSO is 1944, the 53 years processed so far (1964–2016) are extended by the results for solar cycle No. 19 (present paper). The completion of the user-friendly catalog of KSO sunspot group positions, rotation velocities and differential rotation parameters is therefore very important for the further long-term analysis of photospheric differential rotation.

The comparison of the differential rotation parameters obtained from different sources (e.g. KSO, YNAO, GPR, KoSO) for solar cycle No. 19 showed that our results for the equatorial rotation velocity (parameter $A$) and the gradient of differential rotation (parameter $B$) significantly coincide with earlier results from the KSO data (performed with a different method), as well as with results from the KoSO and YNAO data. In contrast, the values of parameter $A$ from three different papers based on the GPR data show statistically significant differences compared to all results from the aforementioned observatories. Although the discrepancies between GPR and KSO have previously been entirely attributed to instrumental origin, and numerical corrections to the KSO results from \citet{lustig1983} were proposed, it seems unlikely that the same systematic errors would also be present in YNAO and KoSO data for cycle No. 19. Since the three independent data sets match and clash with the GPR data set, the findings suggest that the GPR data appear to be inconsistent. Therefore, the discrepancies found between the GPR and others (KSO, YNAO, KoSO) should be thoroughly analysed in the future to identify their possible causes. Furthermore, as a next step, we plan to extend the analysis of the comparison to other available solar cycles.  

The analysis of the north-south asymmetry in the solar rotation profile during solar cycle No. 19 (using two different comparison methods) showed  that the differential rotation parameters of the hemispheres significantly coincide indicating symmetry in the rotation of the Sun around the equator during cycle No. 19. The same is valid for the KSO data of \citet{lustig1983} as well as for the YNAO data. Nevertheless, it is interesting that all these sources show slightly higher equatorial rotation velocities in the southern hemisphere compared to the northern hemisphere.

After completion of the KSO catalog of positions and rotation velocities, a detailed analysis of the north-south asymmetry in the solar rotation profile and its comparison with the asymmetry of activity is planned. This could help to explain the connection between the pronounced north-south asymmetry in the activity and the symmetry in the solar rotation observed for solar cycle No. 19.

%

%
\begin{acks}
This work was partially supported by the University of Rijeka through the project Uniri-iskusni-23-256. Funding was also received from the Horizon 2020 project SOLARNET (824135, 2019–2023). In addition, we recognize the support provided by the Austrian-Croatian Bilateral Scientific Projects ``Comparison of ALMA observations with MHD simulations of coronal waves interacting with coronal holes'', ``Multi-Wavelength Analysis of Solar Rotation Profil'' and ``Analysis of solar eruptive phenomena from cradle to grave''.
\end{acks}

 \begin{authorcontribution}

IPB, RB, RJŠ, AMV and AH contributed to the conception and design of the work. LŠ, IPB, WP and KL contributed to the acquisition and analysis of the data, while IPB, TJ, LŠ and RB were involved in the interpretation of the observations. LŠ, IPB and KL prepared the figures. DH created and edited the Sungrabber software according to the requirements. IPB wrote the first draft of the manuscript, which was critically commented by all other authors. After review, all authors approved the final manuscript.

 \end{authorcontribution}
%
%
 \begin{dataavailability}
The corresponding author can provide the datasets produced and/or analyzed in the current study upon reasonable request.
 \end{dataavailability}
 \begin{ethics}
 \begin{conflict}
No conflicts of interest are declared by the authors.
 \end{conflict}
 \end{ethics}

  
\bibliographystyle{spr-mp-sola}
\bibliography{RefSC19}

\end{document}